\documentclass[twocolumn]{aastex62}

\turnoffedits
\usepackage{amsmath}
\usepackage{stix}
\usepackage{enumitem}
\usepackage{xspace}
\newcommand{\teff}{\ensuremath{T_{\mathrm{eff}}}}
\newcommand{\logg}{\ensuremath{\log g}}

\newcommand{\vsini}{\ensuremath{v \sin i}\xspace}
\newcommand{\mjup}{\ensuremath{M_\mathrm{Jup}}\xspace}

\received{}
\revised{}
\accepted{}

\shorttitle{Spectral Variability of VHS1256\MakeLowercase{b}}
\shortauthors{Zhou, Bowler et al.}

\begin{document}

\title{Spectral Variability of VHS J1256--1257b from 1 to 5 \micron{}}

\begin{abstract}
  Multiwavelength time-resolved observations of rotationally modulated variability from brown dwarfs and giant exoplanets are the most effective method for constraining their heterogeneous atmospheric structures. In a companion paper \citep{Bowler2020}, we reported the discovery of strong near-infrared variability in \textit{HST}/WFC3/G141 light curves of the very red L-dwarf companion VHS J1256--1257b. In this paper, we present a follow-up 36 hr \textit{Spitzer}/IRAC Channel~2 light curve together with an in-depth analysis of the \textit{HST} and the \textit{Spitzer} data. The combined dataset provides time-resolved light curves of VHS1256b sampling 1.1 to 4.5 \micron{}. The \textit{Spitzer} light curve is best fit with a single sine wave with a period of $22.04\pm0.05$~hr and a peak-to-peak amplitude of $5.76\pm0.04$\%. Combining the period with a previously measured projected rotational velocity ($v\sin i$), we find that VHS1256b is most consistent with equatorial viewing geometry. The \textit{HST}/G141+\textit{Spitzer} spectral energy distribution favors a \teff{} of 1000~K, low surface gravity model with disequilibrium chemistry. The spectral variability of VHS1256b is consistent with predictions from partly cloudy models, suggesting heterogeneous clouds are the dominant source of the observed modulations. We find evidence at the $3.3\sigma$ level for amplitude variations within the 1.67\micron{} CH$_{4}$ band, which is the first such detection for a variable L-dwarf. We compare the \textit{HST}/G141  time-resolved spectra of three red L-dwarfs with high-amplitude near-infrared rotational modulations and find that although their time-averaged spectra are similar, their spectroscopic variabilities exhibit notable differences. This diversity reinforces the advantage of time-resolved spectroscopic observations for understanding the atmospheres of brown dwarfs and directly imaged exoplanets.
\end{abstract}

\correspondingauthor{Yifan Zhou}
\email{yifan.zhou@utexas.edu}
\author[0000-0003-2969-6040]{Yifan Zhou}
\altaffiliation{Harlan J. Smith McDonald Observatory Fellow}
\affiliation{Department of Astronomy, The University of Texas at Austin, Austin, TX 78712, USA}
\affiliation{McDonald Observatory, The University of Texas at Austin, Austin, TX 78712, USA}
\author{Brendan P. Bowler}
\affiliation{Department of Astronomy, The University of Texas at Austin, Austin, TX 78712, USA}
\author{Caroline V. Morley}
\affiliation{Department of Astronomy, The University of Texas at Austin, Austin, TX 78712, USA}
\author{D\'aniel Apai}
\affiliation{Department of Astronomy/Steward Observatory, The University of Arizona, 933 N. Cherry Avenue, Tucson, AZ, 85721, USA}
\affiliation{Department of Planetary Science/Lunar and Planetary Laboratory, The University of Arizona, 1640 E. University Boulevard, Tucson, AZ, 85718, USA}
\affiliation{Earths in Other Solar Systems Team, NASA Nexus for Exoplanet System Science.}
\author{Tiffany Kataria}
\affiliation{Jet Propulsion Laboratory, California Institute of Technology, 4800 Oak Grove Drive, Pasadena, CA, USA}
\author{Marta L. Bryan}
\affiliation{Department of Astronomy,  University of California Berkeley, Berkeley, CA 94720-3411, USA}
\author{Bj\"orn Benneke}
\affiliation{University of Montreal, Montreal, QC, H3T 1J4, Canada}

\renewcommand{\plotone}[1]{\includegraphics[width=\columnwidth]{#1}}

\section{Introduction}

The formation of condensate clouds is a signature characteristic of the atmospheres of brown dwarfs \citep[e.g.,][]{Kirkpatrick2005,Marley2015} and exoplanets \citep[e.g.,][]{Madhusudhan2019}. The formation and dissipation of clouds with changing effective temperatures are recognized as the primary driver of spectral-type and color evolution for brown dwarfs \citep[e.g.,][]{Knapp2004,2008ApJ...678.1372C,Saumon2008}. Thick clouds due to low surface gravity are attributed as the cause of the extremely red near-infrared colors seen in young directly imaged planetary-mass objects and brown dwarfs near the L/T transition \citep[e.g.,][]{Liu2016}. Strong cloud opacity also obscures substellar atmospheres and introduces significant uncertainty in the retrieval analysis of atmospheric compositions \citep[e.g.,][]{Burningham2017a}. \edit1{It is notable that studies \citep[e.g.,][]{Tremblin2016,Tremblin2017} were put forward to explain these observations without the introduction of clouds. Nevertheless, they do not argue against the significance of clouds in understanding the substellar atmospheres.} Properly measuring and modeling cloud structures is critical for characterize brown dwarfs and exoplanets.

Time-series observations of rotational modulations in ultracool brown dwarfs have deepened our understanding of condensate clouds. The most likely explanation for the observed rotational modulation is heterogeneous clouds \citep[e.g.,][]{Apai2013,Morley2014}, which is supported by multiple time-series observational surveys \citep[e.g.,][]{Radigan2012,Buenzli2014,Metchev2015}. These studies found that while rotational modulations are observed across almost the entire range of spectral types of ultracool dwarfs, brown dwarfs in the L/T transition have a significantly higher probability of exhibiting high-amplitude variability. This finding agrees with the picture that silicate and iron clouds begin to submerge below the photosphere of brown dwarfs in the L/T transition \citep[e.g.,][]{Saumon2008, Marley2010}. Consequently, the heterogeneous cloud top causes strong rotationally modulated variability of the hemispherically-integrated light.

High-amplitude (peak-to-peak ${>}2\%$) variable brown dwarfs discovered in these surveys further motivated time-resolved spectroscopic observations \citep[e.g.,][]{Apai2013,Buenzli2015,Biller2017,Zhou2018}. The chromatic variations in the amplitude and the phase of rotational modulations can constrain vertical cloud structures. For example, because the 1.4 \micron{} region probes relatively high altitude atmospheres whereas the neighboring continuum probes lower altitudes,  modulation amplitudes decrease at the 1.4 \micron{} water absorption band when patchy clouds are at intermediate altitudes. These wavelength-dependent modulations have been observed in several brown dwarfs at the L/T transition \citep[e.g.,][]{Apai2013,Yang2014,Zhou2018}. For several mid-L type brown dwarfs, the amplitudes of chromatic variations are not nearly as strong \citep{Yang2014, Yang2016}, suggesting that cloud layers are above the pressure levels where the optical depths equal to one ($\tau=1$) both in the 1.4 \micron{} water absorption band and in the continuum. These results generally fit into the picture of silicate and iron condensate clouds descending in the atmosphere with decreasing effective temperatures. Recent observations of very red variable late-L brown dwarfs showed that the wavelength dependence of these objects are in between mid-L dwarfs and the L/T transition dwarfs \citep{Lew2016, Biller2017}. These new findings raise questions, such as whether cloud evolution naturally explains their chromatic rotational modulations, or if the onset of the L/T transition occurs in the same way with these reddened brown dwarfs as it does in regular field brown dwarfs. Because these late-L type brown dwarfs have very similar spectra to directly imaged giant exoplanets such as HR8799bcde \citep{Marois2008a,Marois2010} and 2M1207b \citep{Chauvin2004}, time-series observational studies of red L-type brown dwarfs can further our understanding of the atmospheres of these exoplanets.

Time-domain monitoring also reveals atmospheric dynamic processes, circulation patterns, and weather in substellar atmospheres \citep[e.g.,][]{Apai2017}. Based on general circulation models, cloud coverage is expected to evolve on timescales as short as a few rotation periods for brown dwarfs \citep[e.g.,][]{Showman2012,Zhang2014,Tan2018}. Indeed, \citet{Apai2017} found significant light curve evolution in multiple brown dwarfs. These authors proposed that planetary-scale waves modulate the scale height of the clouds are the most likely cause. Long time-baseline multiwavelength observations, which probe the atmospheric evolution at different pressure levels, are the most direct and effective method for detecting these atmospheric activities.

The rotation rate is another fundamental property that can be measured from time-resolved observations. The derived angular momentum can then be used to trace the formation and dynamical evolution of substellar systems \citep[e.g.,][]{Bryan2018, Scholz2018}.   The orientation angle of the spin axis can be inferred by combining the rotation period and the projected spin velocity ($v \sin i$). \edit1{Because formation and dynamic evolution leave traces in the alignment of the angular momentum vectors of the components of the systems, the orientation of the spin axis in binary and planetary systems provides information about the formation and angular momentum history of brown dwarfs and planets \citep[e.g.,][]{Bowler2017, Bonnefoy2018, Bryan2018, Bryan2020, Xuan2020}}. Additionally, the observed modulation amplitude and near-infrared colors of brown dwarfs have shown to be correlated with their spin axis inclination \citep[e.g.,][]{Vos2017, Vos2018a}. 

\object{VHS J125060.192-125723.9b} (hereinafter VHS1256b, discovered by \citealt{Gauza2015})  has emerged as a prime target for studying the atmospheric properties of directly imaged exoplanets. With an estimated mass of 10-25 \mjup{} \citep{Gauza2015,Rich2016}, VHS1256b is on the planet-brown dwarf boundary $({\sim}13\mjup)$. It orbits its host \citep{Rich2016,Stone2016}, a late-M binary (\edit1{123.6 mas angular separation}) brown dwarfs, at a projected separation of $105$ au. Thus, VHS1256b and its host binary form a rare hierarchical triple system that has the potential to reveal crucial information on the formation and dynamical evolution of similar systems. Spectroscopic observations \citep{Gauza2015, Rich2016} found that the  near-infrared color and spectrum (L7 spectral type) of VHS1256b are remarkably similar to the free-floating planetary-mass objects PSOJ318 \citep{Liu2013,Biller2017} and WISEJ0047 \citep{Gizis2012,Lew2016} as well as the exoplanets HR8799bcde \citep{Marois2008a,Marois2010}. The wide separation  (8\arcsec.1 angular separation) and low contrast ($\Delta J$=5.6~mag) between VHS1256b and its host star make it much more favorable for high-precision follow-up characterization than the HR8799 planets. In a companion paper, \citet{Bowler2020} reported ${\sim}20\%$ overall variability in the \textit{Hubble Space Telescope}/Wide Field Camera 3 (\textit{HST}/WFC3) G141 grism  bandpass ($1.1$--$1.7$\micron) over the course of six consecutive \textit{HST} orbits. The variability at 1.27~\micron{} (the peak of $J$-band) is nearly 25\%, which is the second-highest variability amplitude ever recorded in a substellar object. This high variability amplitude makes VHS1256b an ideal target for characterizing patchy clouds in planetary atmospheres.

The observations presented in \citet{Bowler2020} did not cover a full rotation period of VHS1256b. Therefore, \citet{Bowler2020} only placed lower limits on the modulation amplitude and a rough constraint on the rotation period. The estimate of the rotation period was also strongly affected by the assumed light curve model. In this paper, we first present a  new 4.5\micron{} \textit{Spitzer} light curve of VHS1256b and precisely determine its rotation period. We then combine the results from the \textit{Spitzer} light curve and the \textit{HST}/WFC3 time-resolved spectroscopy to investigate in detail the chromatic variations, which span 1.1-4.5\micron{}.  We also place this object  in the context of very red variable L-dwarfs and discuss the implications of its variability.

\section{Observations}

\begin{figure}
  \centering
  \plotone{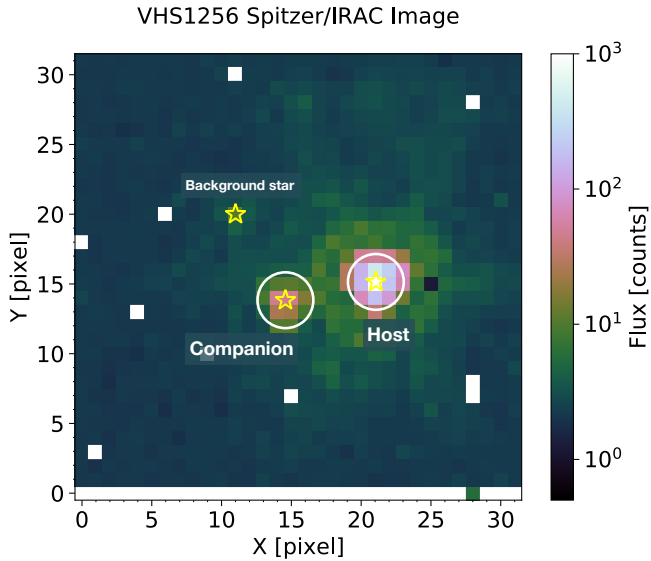}
  \vspace{-5em}
  \caption{An example \textit{Spitzer}/IRAC Channel 2 image of the VHS1256 system. The companion (VHS1256b), the host binary (VHS1256AB, \edit1{unresolved}), and a faint background star (2MASS J1256019--1257390) are visible. White circles mark the apertures adopted to measure the photometry for each component. Pixels overlaid by white masks are bad pixels identified by the IRAC Level 1 pipeline. We exclude these pixels during the data reduction.}
  \label{fig:SpitzerImage}
\end{figure}
We observed VHS1256b with the \textit{Spitzer Space Telescope}'s InfraRed Array Camera \citep[IRAC,][]{Fazio2004} in Channel 2 (4.5\micron{})  consecutively for 37.1~hr from 2019-11-14 05:45 to 2019-11-15 18:17 UTC  (\textit{Spitzer} DDT program 14312, PI: Zhou). The observations consisted of five astronomical observation requests (AORs). The first and last AORs were short (30~min and 10~min, respectively); we used these two AORs to measure the sky darks by pointing the telescope to empty patches of sky that were a few arcminutes away from our target. The second to the fourth AORs (``scientific AORs'') were all 12~hr long. Together, they formed a 36~hr continuous on-target monitoring sequence.

Instrument setups were kept identical throughout the entire scientific observations. We planned our observation following the recommendations for high-precision photometry observations provided by the \emph{Spitzer Science Center} (SSC)\footnote{The recommendations for optimizing analysis techniques for exoplanet and brown dwarf light curve studies can be found at \url{https://irachpp.spitzer.caltech.edu}.}. Procedures were implemented to place VHS1256b on the ``sweet spot'' of the detector. However, the actual telescope pointing deviated from the requested position by about half a pixel in both $x$ and $y$ directions, which limited the utility of the IRAC center pixel response map \citep{Ingalls2012} to correct light curve systematics. We adopted the $32\times32$ subarray that corresponded to a field of view of $38\arcsec\times38\arcsec$. To prevent the bright host binary from saturation and facilitate systematics correction, we adopted a high frame rate with an exposure time of 2~s. In total, we obtained 63,744 images from the three scientific AORs. In the \emph{Spitzer} subarray mode, every 64 frames taken in succession are stored in one single-header FITS file. Therefore, our scientific observations resulted in 996 FITS files. Figure \ref{fig:SpitzerImage} shows one example image from the observations.

We also make use of the data from \textit{HST} program GO-15197 (PI: Bowler) observed from 2018-03-05 to 2018-03-06 UT. These observations consist of WFC3/G141 spectroscopy spanning 1.1 to 1.7 \micron{} in six contiguous orbits and were first described in our companion paper (Bowler et al. 2020). In this study, we jointly analyze the \textit{Spitzer} and \textit{HST} observations.

\section{Data Reduction}
In this section, we describe the procedures to extract and calibrate light curves from the  \textit{Spitzer} Channel 2 observations.
The SSC reduced the raw data using the IRAC Level 1 pipeline (version: S19.2.0) and delivered the basic calibrated data (\texttt{BCD}) files. The IRAC pipeline reduction included instrument signature corrections, scattered light, dark current removal, and flat-field calibration. Details of the IRAC pipeline processes are given in the  \href{https://irsa.ipac.caltech.edu/data/SPITZER/docs/irac/iracinstrumenthandbook/}{IRAC Instrument Handbook}. Each of the \texttt{BCD} file contained a single header and 64 reduced images. Our reduction started with the \texttt{BCD} files.

\subsection{Preparations}

\begin{figure*}
  \centering
  \plottwo{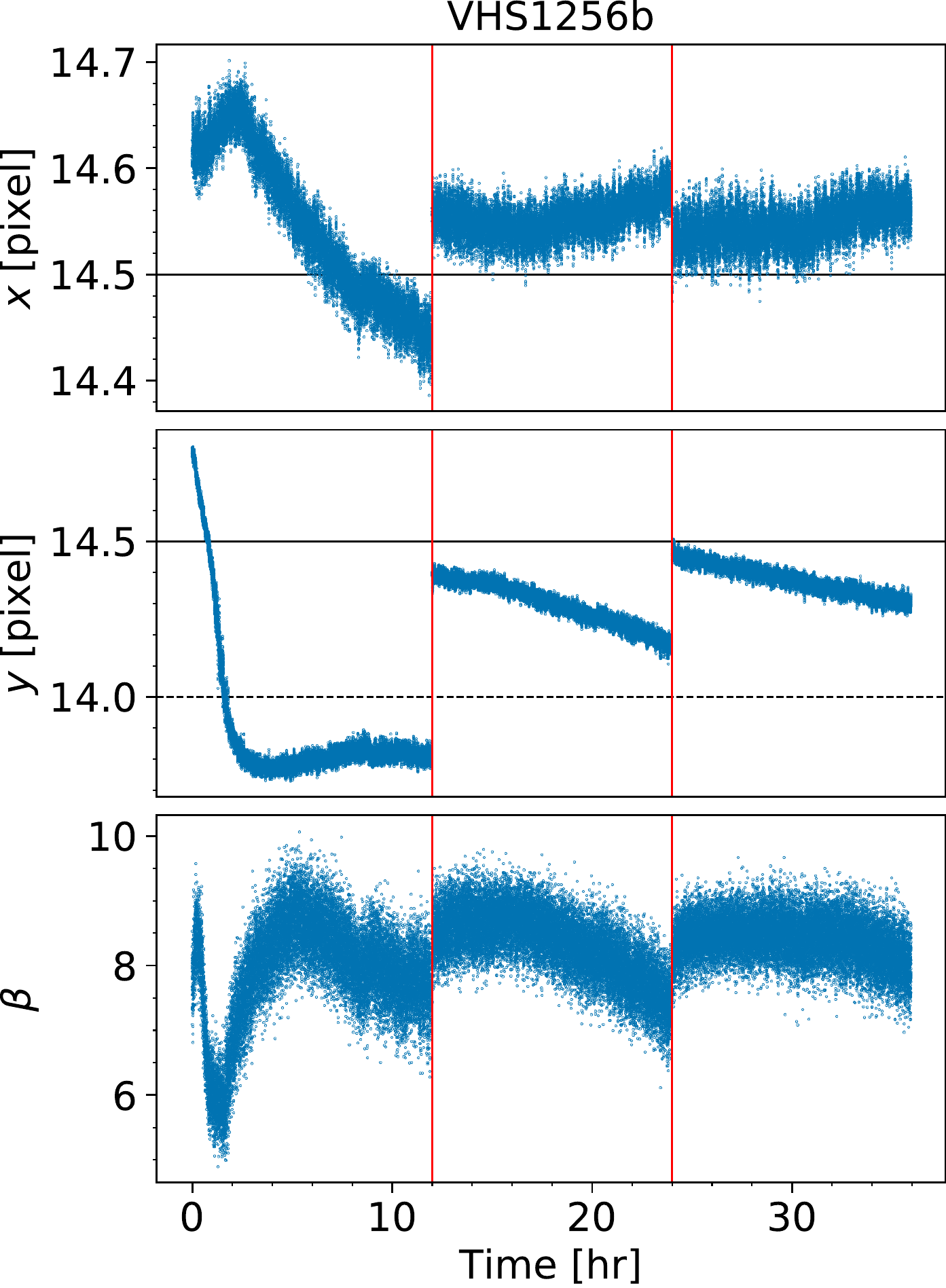}{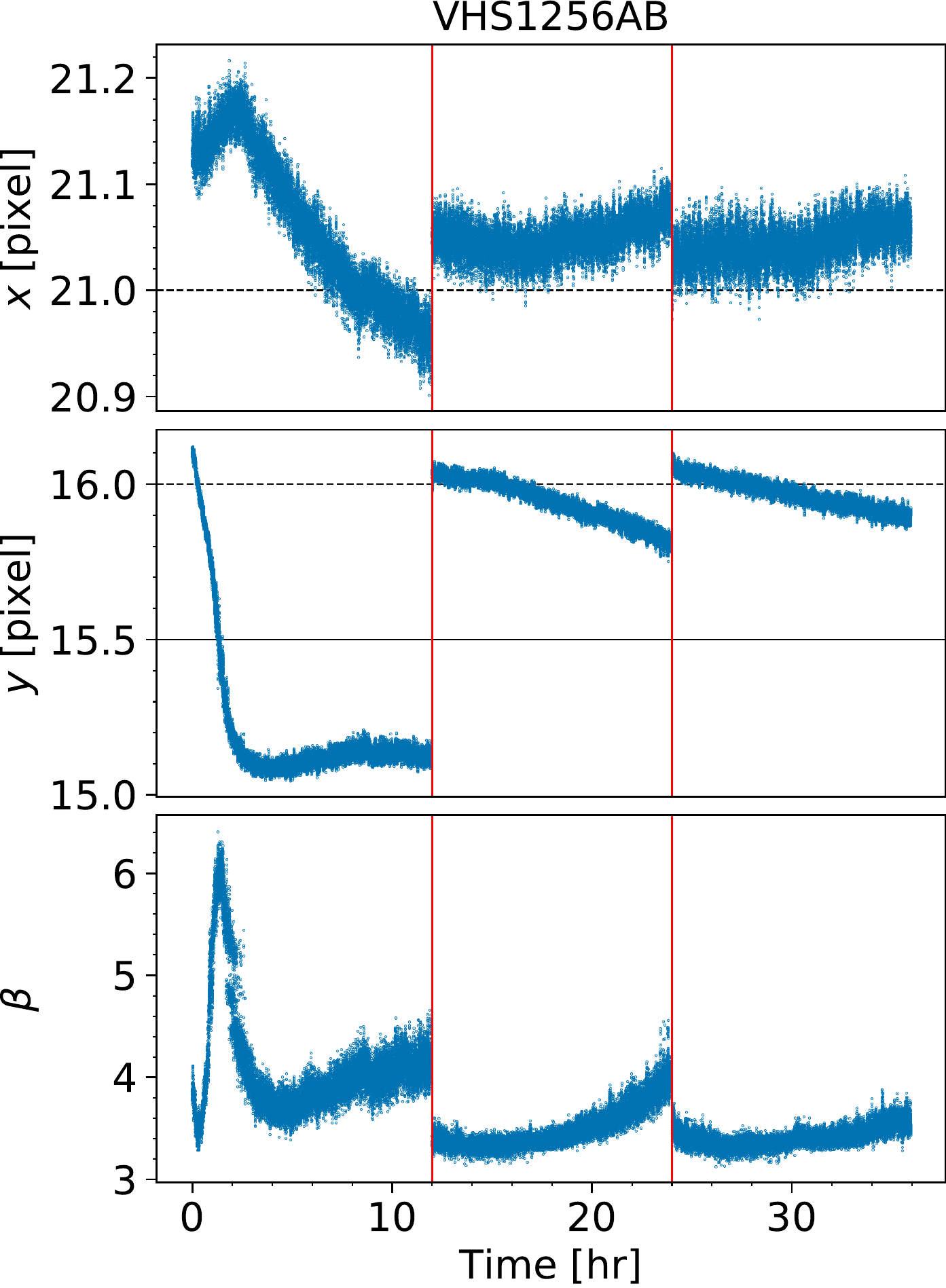}
  \caption{\textit{Spitzer} diagnostic trends for VHS1256b (left) and the host binary (right). The top and middle panels show the centroid drifts in the $x$ and $y$ directions. Pixel centers and edges are indicated by dashed and solid lines, respectively. The bottom panels show the trends of noise pixel values ($\beta$ defined in Equation~\ref{eq:NP}). We later used these trends to remove systematics from our light curves.}
  \label{fig:trends}
\end{figure*}

First, we assigned a timestamp to each image. Because the FITS file header only has one timestamp per 64 images, we interpolated the remaining timestamps. We assumed that the 64 images in each FITS file had an identical frame time and calculated their timestamps by evenly interpolating between the values of header keywords \texttt{AINTBEG} (the beginning of the integration)  and \texttt{ATIMEEND} (the end of the integration). We then sorted all images chronologically and placed them into a single data cube.

Next, we searched and corrected for bad pixels and cosmic rays. In the IRAC Level 1 pipeline reduction, bad pixels were flagged and stored in the \texttt{bimsk} files, which were delivered along with the \texttt{BCD} files. As shown in Figure~\ref{fig:SpitzerImage}, none of the pipeline-identified bad pixels were close to either the companion or the host. Therefore, we simply excluded these pixels from our subsequent analyses. To find cosmic rays, we computed a median-filtered light curve for each pixel (filter width: 33 frames) and compared it with the original one. $5\sigma$ outliers were identified as cosmic rays, and the values of these pixels were replaced with linear interpolations of pixel values in the neighboring frames.

We then measured the centroids for both the companion and the host by fitting 2D Gaussian profiles to each image. We did not adopt the image-moment-based method that is used in the \texttt{box\_centroid.pro} code provided by the SSC because potential contamination from the bright host could bias the image-moment calculations of the fainter companion, whereas this contamination had negligible effect on 2D Gaussian fittings. We found that the centroid movement trends of the companion and the host were consistent with each other. However, the scatter in the primary's centroid measurements was much smaller because the signal-to-noise ratio (SNR) of its image is higher. We therefore adopted relative centroid changes of the host ($\Delta x_{\mathrm{host}} = x_{\mathrm{host}}-\mathrm{median(x_{\mathrm{host}})}, \Delta y_{\mathrm{host}} = y_{\mathrm{host}}-\mathrm{median(y_{\mathrm{host}})}$) for the telescope-pointing motions. For the companion, we first calculated the median of the $x$ and $y$ measurements ($x_{\mathrm{companion}}, y_{\mathrm{companion}}$) and then reconstructed the centroid time-series by adding $\Delta x_{\mathrm{host}}$ and $\Delta y_{\mathrm{host}}$ to $x_{\mathrm{companion}}$ and $y_{\mathrm{companion}}$, respectively.  As shown in  Figure~\ref{fig:trends}, the centroid trends reveal that while the telescope-pointing position was relatively stable in the second and third scientific AORs, it had significant drift in the first two hours of the first scientific AOR.

For the sky background, we optimized estimation procedures for the photometry of the host and the companion. For the host, we adopted a ``global sky'' value, which was estimated using pixels that were not significantly illuminated by any astrophysical sources. Specifically, we first masked out pixels that were within an 8-pixel radius of the host, a 5-pixel radius of the companion, and a 5-pixel radius of the background star. We then conducted a $2\sigma$ ten-iteration sigma-clipping on the unmasked part of the image to further exclude any illuminated pixels. The median value of the remaining unmasked pixels was adopted as the ``global sky'' value. For the companion, the sky background estimation took into account possible contamination of the host star. We opted for a ``local sky'' value: a sky background  was estimated using pixels that were in an annulus spanning five to eight pixels away from the centroid of the companion. Pixels that were within five pixels from the host or 2.5 pixels from the background stars were excluded. The median values of pixels in this annulus was adopted as the ``local sky'' value.

\subsection{Photometry}

We conducted aperture photometry with an aperture size of 2 pixels in radius (Figure~\ref{fig:SpitzerImage}) for both the companion and the host. Aperture photometry was implemented using  the \texttt{aperture\_photometry} function in the Python package \texttt{photutils} \citep{Bradley2019}. We executed this function with the option \texttt{method=`exact'} so that the exact intersections of the aperture and pixels were used to determine the proportional flux for pixels that were partially included in the aperture. ``Local sky'' and ``global sky'' values were subtracted from these measurements for the companion and the host, respectively.  The initial photometry results are shown in Figure~\ref{fig:rawlc}. They exhibit clear systematic trends.

We calculated the ``noise pixel'' parameter ($\beta$, IRAC Handbook \S 2.2.2, also see e.g., \citealt{Lewis2013}) using the same aperture as we used for the  photometry. This parameter, defined in Equation~\ref{eq:NP}, quantifies the sharpness of the image and is useful in decorrelating the systematic trends from \emph{Spitzer}/IRAC light curves \citep[e.g.,][]{Knutson2012,Lewis2013},
\begin{equation}
  \label{eq:NP}
  \beta = \frac{\Bigl(\sum P_{i} \Bigr)^{2}}{\sum P_{i}^{2}}.
\end{equation}
Here, $P_{i}$ represents the pixel response function (PRF) and the sum operators total the PRFs or their squares within an aperture. In practice, $P_{i}$ can be replaced by the pixel value $F_{i}$ in the observations. We used the same method as in aperture photometry to derive the noise pixel values for both the companion and the host. The noise pixel time-series plots are presented in Figure~\ref{fig:trends}. There are correlated variations between the trends of the noise pixels and the centroids for both components of the VHS1256 system.

\subsection{Systematic Correction}

\begin{figure*}[!t]
  \centering
  \plottwo{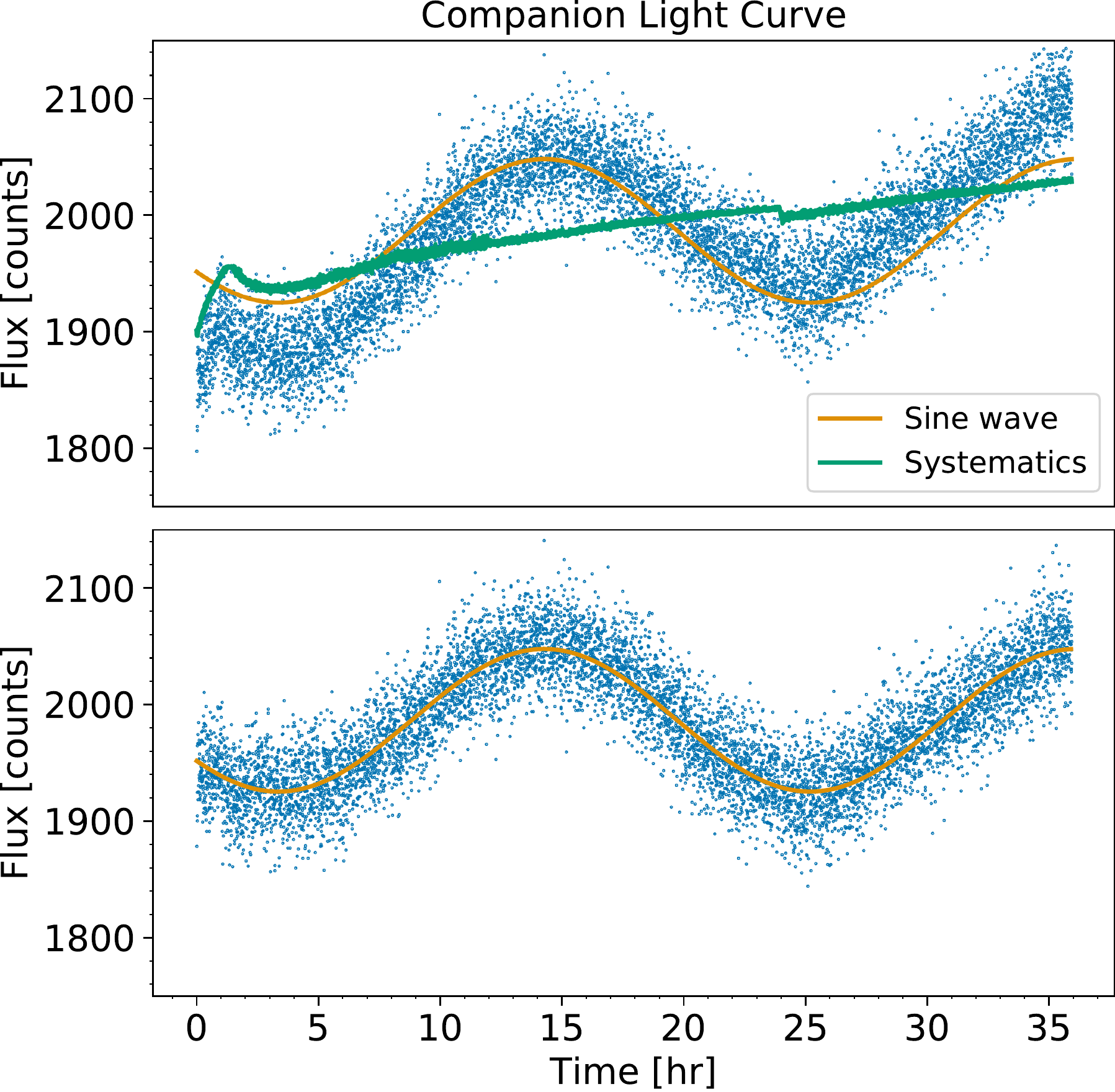}{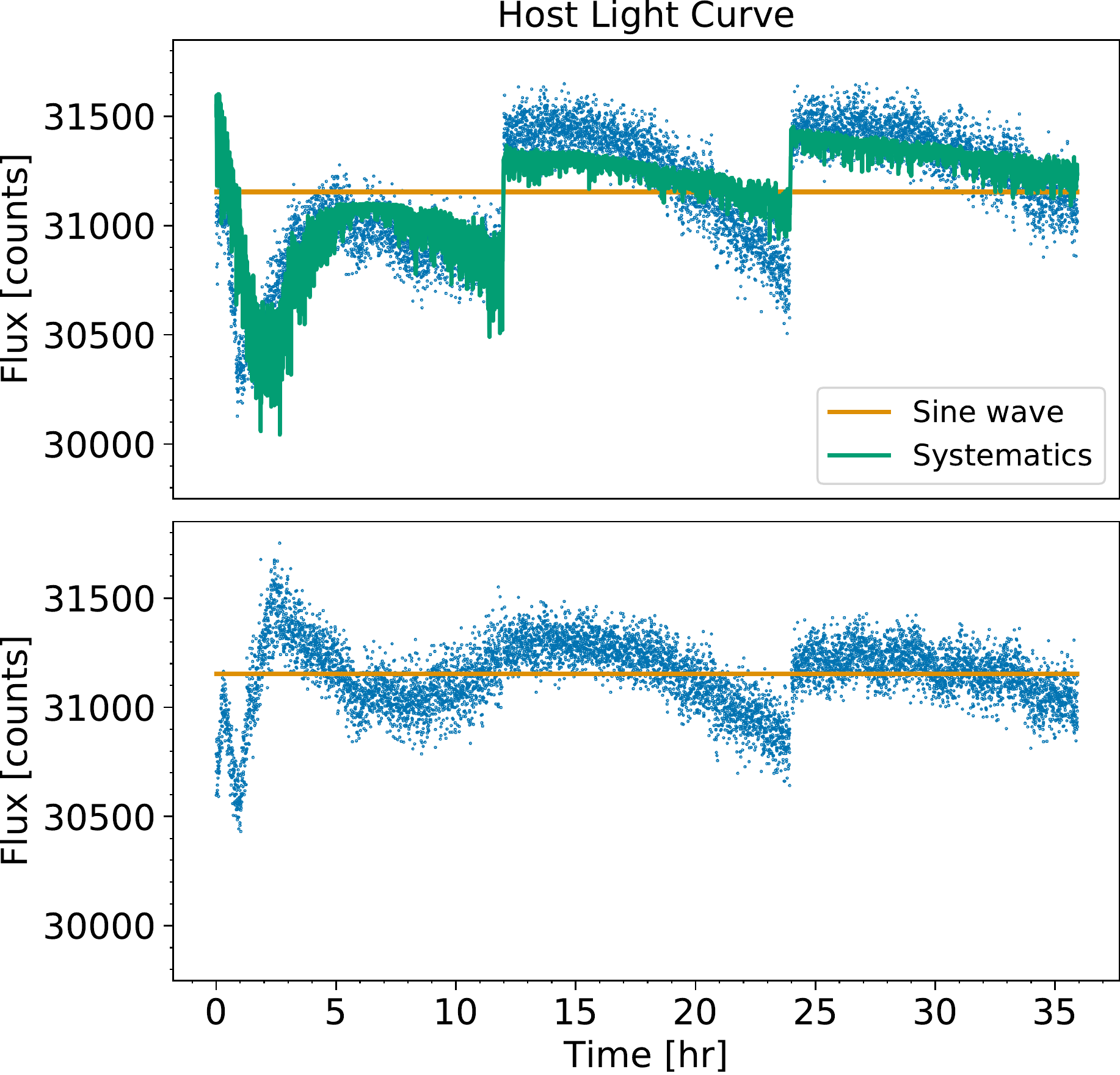}
  \caption{Raw aperture-photometry light curves of VHS1256b (left) and the host binary (right) and the results from polynomial method corrections. The raw light curves (upper panel)  demonstrate systematic noise that is correlated with the \textit{Spitzer} diagnostic trends (Figure~\ref{fig:trends}). After it is divided by the systematic models constructed using Equation~\ref{eq:poly}, the companion light curve agrees almost perfectly with a  sine wave. However, strong systematic trends remain in the host light curve. The imperfect correction for the host light curve calls for more advanced correction methods.}
  \label{fig:rawlc}
\end{figure*}

Intra-/inter-pixel-sensitivity variations coupled with telescope-pointing drifts often produce notable systematic trends in \textit{Spitzer}/IRAC light curves. These systematics manifest as discontinuities between AORs and correlated variations between light curves and telescope pointing positions and are clearly visible in our raw light curves (Figure~\ref{fig:rawlc}). Various approaches to correct for these trends have been developed for transiting exoplanets  \citep[e.g.,][]{Lewis2013,Deming2014,Kilpatrick2016,Dang2018} and variable brown dwarf observations \citep[e.g.,][]{Heinze2013,Metchev2015,Yang2016}. In the following, we implement two methods and determine the most suitable approach for removing systematics in our light curves.

\subsubsection{The Polynomial Method}

Our first approach is to construct polynomial fits using telescope-pointing motion ($x$, $y$) to decorrelate systematic trends from the astrophysical signals. These methods were applied in various brown dwarf light curve studies \citep[e.g.,][]{Metchev2015,Yang2016} as well as hot Jupiter phase curve analyses \citep[e.g.,][]{Dang2018}. We adopted the second-order polynomials of the form
\begin{equation}
  \label{eq:poly}
  \mathrm{sys_{poly}} = 1 + p_{1,1}\Delta x + p_{1,2}\Delta y + p_{2,1}\Delta x^{2} + p_{2,2}\Delta y^{2}
\end{equation}
to model the the intra-/inter-pixel related trends.
In Equation~\ref{eq:poly}, the coefficients $p_{i,j}$ are free parameters. They are optimized to minimize fitting residuals. In addition to Equation~\ref{eq:poly}, we included a linear function of time as part of the systematic model.

The results are presented in Figure~\ref{fig:rawlc}. For the companion light curve, this method works well. The corrected light curve shows no apparent discontinuities or correlated noise. However, for the host, strong systematic trends remain after the polynomial correction, indicating that the polynomial method was inadequate. Therefore we also implemented more advanced methods for systematics correction. 

\subsubsection{Pixel-mapping Method}

\begin{figure*}[!t]
  \centering
  \includegraphics[width=0.48\textwidth]{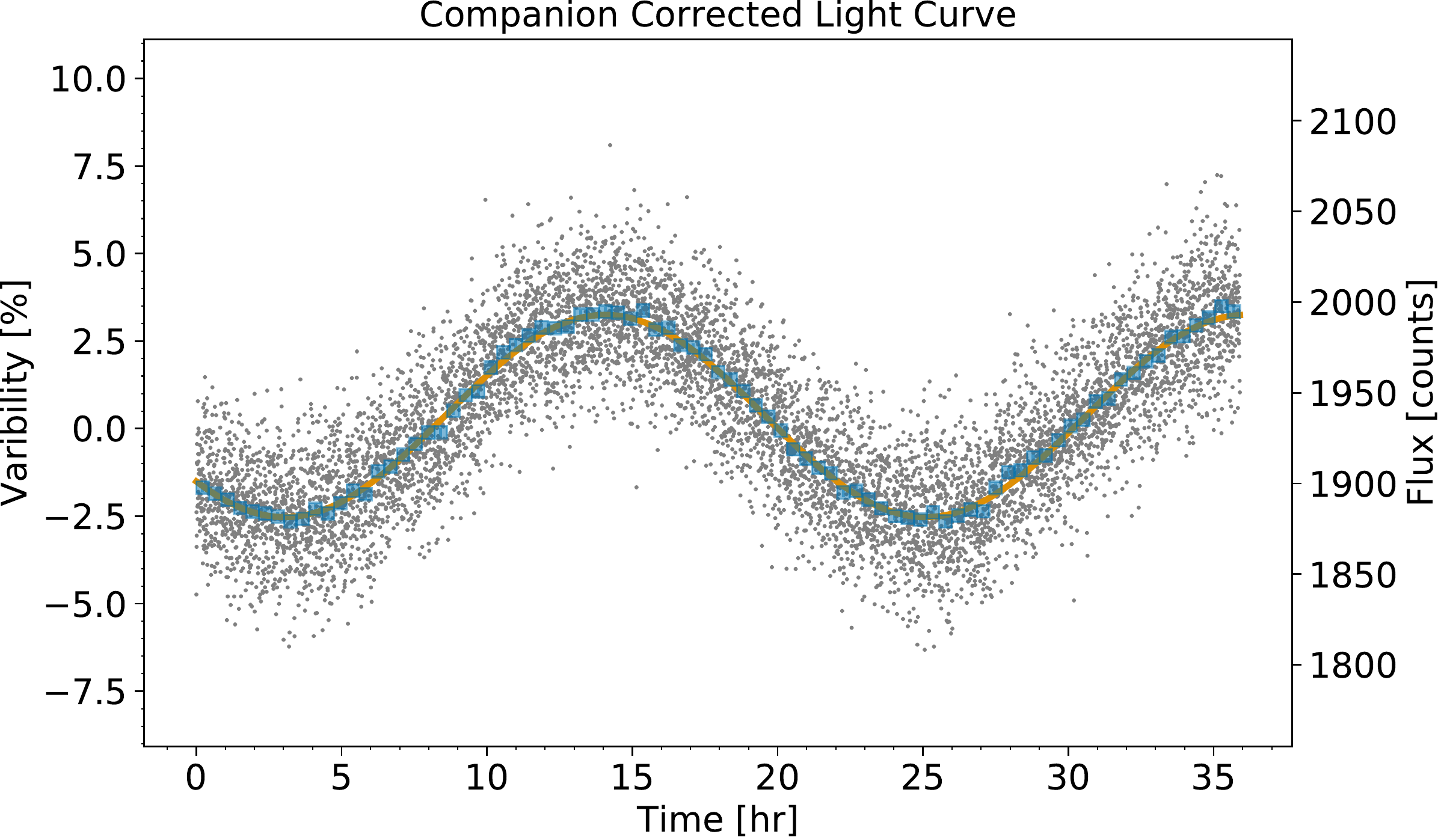}  
  \includegraphics[width=0.48\textwidth]{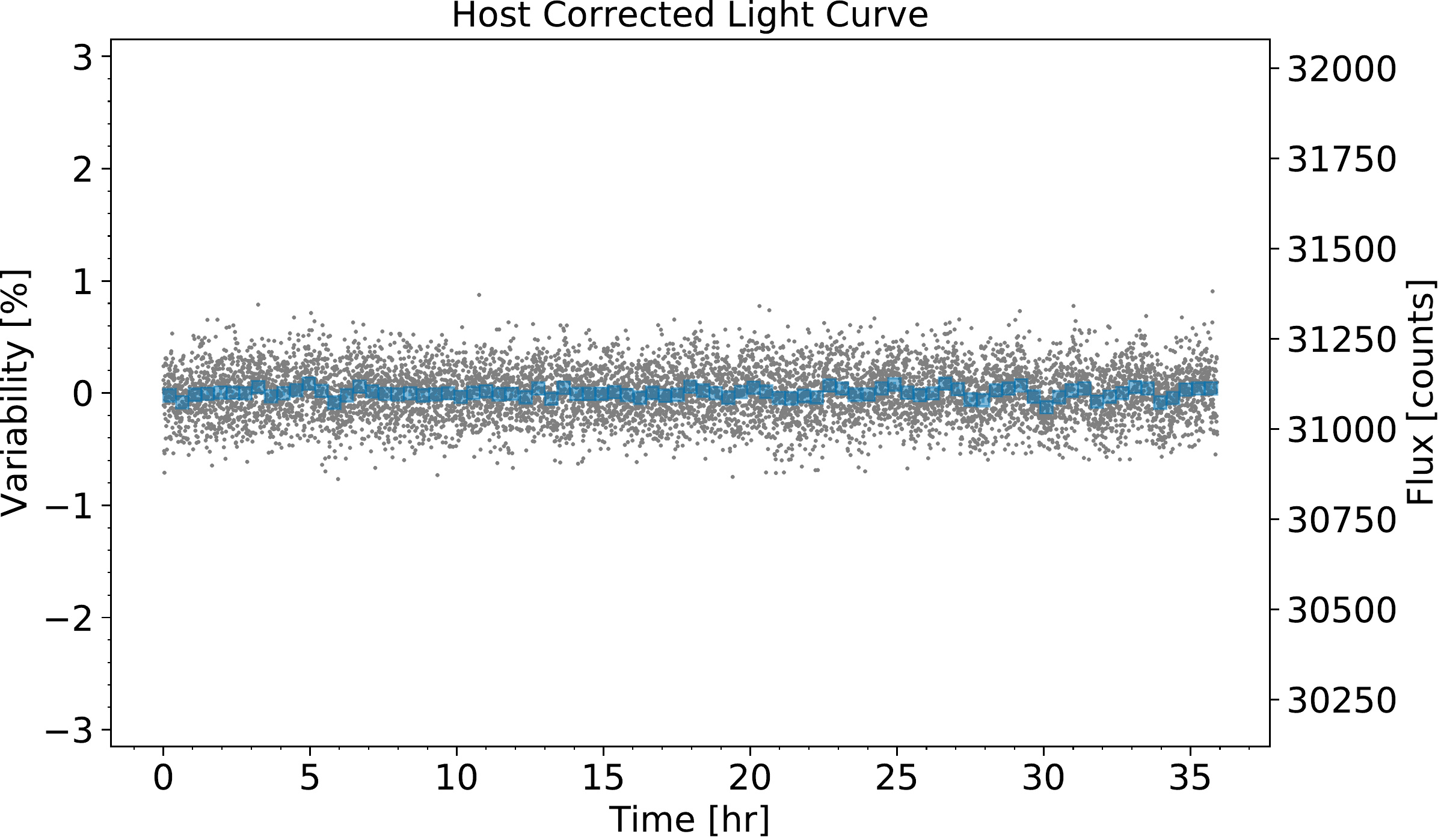}
  \caption{Light curves of VHS1256b and VHS1256AB after pixel-mapping corrections. Systematic trends presented in Figure~\ref{fig:rawlc} are eliminated for both the companion and the host. Although we assume single sine wave models for both objects at the beginning of the iterative correction, their final corrected light curves are different. The corrected light curve of the companion has a sinusoidal modulation with an amplitude of 5.76\%. The corrected light curve of the host agrees with a flat line except for a low-level ($<0.1\%$) periodic variation in the third AOR.}
  \label{fig:correctedlc}
\end{figure*}

We implemented the pixel-mapping method \citep{Ballard2010,Lewis2013} to improve the systematics correction. We chose this method over other popular approaches \citep[e.g.,][]{Deming2014} because the level of telescope-pointing drifts in our observations are similar to  those in \citet{Lewis2013}, for which the pixel-mapping method succeeded in removing systematics, whereas other methods are more suitable when the drift levels are lower \citep{Kilpatrick2016}.

The essence of this method is to reconstruct a pixel-sensitivity map as a function of $x$, $y$, and the square root of the noise pixel parameter $\sqrt{\beta}$ in the form of Equation~\ref{eq:3}:
\newcommand{\sqNP}{\ensuremath{\sqrt{\beta}}}
\begin{equation}
  \label{eq:3}
  W_{i}(x_{i}, y_{i}, \sqNP_{i}) = \frac{\sum_{j\ne i} K(i, j) F_{j}}{\sum_{j\ne i}K(i, j)}.
\end{equation}
Here $W_{i}$ is the sensitivity weight for the $i$th frame,  and $F_{j}$ is the measured flux (assuming the source has a constant flux) in the $j$th frame. $K(i, j)$ is a Gaussian smoothing kernel,
\begin{equation}
  \label{eq:4}
  K(i, j) = \exp\Biggl( -\frac{(x_{i}-x_{j})^{2}}{2\sigma_{x}^{2}}  -\frac{(y_{i}-y_{j})^{2}}{2\sigma_{y}^{2}} - \frac{(\sqNP_{i}-\sqNP_{j})^{2}}{2\sigma_{\sqNP}^{2}} \Biggr)
\end{equation}

As suggested by the form of the smoothing kernel function, the contribution from image $j$ to weight $W_i$ damps quickly as the distance in the $(x, y, \sqNP)$ space increases. Therefore, for any image $i$, images that have a large distance in the $(x, y, \sqNP)$ space can be ignored in calculating $W_{i}$. In practice, we use 50 images to calculate each $W_{i}$. This is similar to the procedure adopted in \citet{Lewis2013}. Importantly, in Equation~\ref{eq:3}, the term $F_{j}$ assumes that the source has constant flux, which was not the case in our observations. Therefore, we have to first divide out the astrophysical models from the light curves before  implementing the pixel-mapping corrections.  We then repeat this procedure to solve the astrophysical model and systematic model iteratively, \edit1{following the fitting strategy detailed in \citep{Lewis2013}.} For both the companion and the host, the astrophysical models are assumed to be sine waves. \edit1{In each iteration step, we first fit a sine wave (using \texttt{lmfit}; \citealt{Newville2014}) to the corrected light curve. We then divide this sine wave out of the corrected light curve to recover the average flux, which is represented by the term $F$. Finally, we insert the recovered flux $F$ to Equation~\ref{eq:4} to update the correction term for the light curve.} \edit1{For both the companion and the host, the fitting residuals stay constant after 50 iteration steps.}

The corrected light curves are shown in Figure~\ref{fig:correctedlc}. For the companion, this method and the polynomial method have almost identical results. The systematic trends are eliminated and the corrected light curve is consistent with a single sine wave. For the host, this method performs much better. The discontinuities and correlated noise are eliminated. The corrected light curve is generally consistent with a flat line\footnote{A low-level (A=$0.074\pm0.007\%$) periodic ($P=2.12\pm0.02$~hr) variation is present in the light curve of the third AOR but is not detected in other parts of the light curve. It is uncelar if its origin is astrophysical or instrumental.}.   We use the results from the pixel-mapping correction in our subsequent analysis.


\section{Results}
\subsection{\textit{Spitzer} Channel 2 Flux Density}
We obtain a time-averaged flux density of $\bar{f}=1937.21\pm0.29$~counts per exposure for VHS1256b (aperture radius$=2.5$ pixels) in the \textit{Spitzer} Channel 2 observations. Applying aperture correction and unit conversion, we find $\bar{f_{\nu}}=1703.56\pm0.26$~$\mu$Jy (12.558 mag with $f_{{\nu,0}}=179.7$~Jy) at 4.493~\micron{} or $\bar{f_{\lambda}}=2.5266\pm0.0004\times10^{-13}$~erg\,cm$^{-2}$\,s$^{-1}\,$\micron$^{-1}$. We note that our \textit{Spitzer} Channel 2 flux density is only about 50\% of  its $M$-band flux \citep{Rich2016}, which has a similar wavelength coverage as the \textit{Spitzer} Channel 2. The source of this inconsistency is unclear but is unlikely to be caused by variability, which would have to be at an unprecedented 50\% level to reconcile the two measurements. We note that our \textit{Spitzer} Channel 2 flux density as part of the spectral energy distribution (SED) of VHS1256b is in excellent agreement with both models and observations of brown dwarfs with similar spectral types to VHS1256b \citep[e.g.,][see discussions in \S\ref{sec:results:model}, \ref{sec:discussion:comparison}]{Biller2017}.

\subsection{Updated Physical Properties for VHS1256b}

\edit1{\citet{Dupuy2020}} recently obtained new parallax measurements for VHS1256b. The updated parallax is $45.0\pm2.4$ mas, corresponding to a distance of $22.2\pm1.2$~pc. This new distance measurement exceeds the original estimate of \citet{Gauza2015} by 75\% and is in better agreement with the system's spectroscopic distance ($17.2\pm2.6$~pc, \citealt{Stone2016}). 

Because of the increase in distance, the bolometric luminosity of VHS1256b  increases from $\log(L/L_{\odot})=-5.05\pm0.22$ to $\log(L/L_{\odot})=-4.54\pm0.07$ \citep{Dupuy2020}. Using the updated parallax, \citet{Dupuy2020} found the mass, effective temperature, and surface gravity of VHS1256b to be $M=19\pm5\mjup$, $\teff=1240\pm50$~K, and $\logg=4.55^{+0.15}_{-0.11}$~dex, respectively. Based on the same analysis, the posterior distribution of the radius ($R_{\mathrm{p}}$)  has a $2\sigma$ range from 1.13 to 1.21 $R_{\mathrm{Jup}}$.


\subsection{Rotational Modulations of VHS1256b at 4.5 \micron{}}

\begin{figure}
  \centering
  \plotone{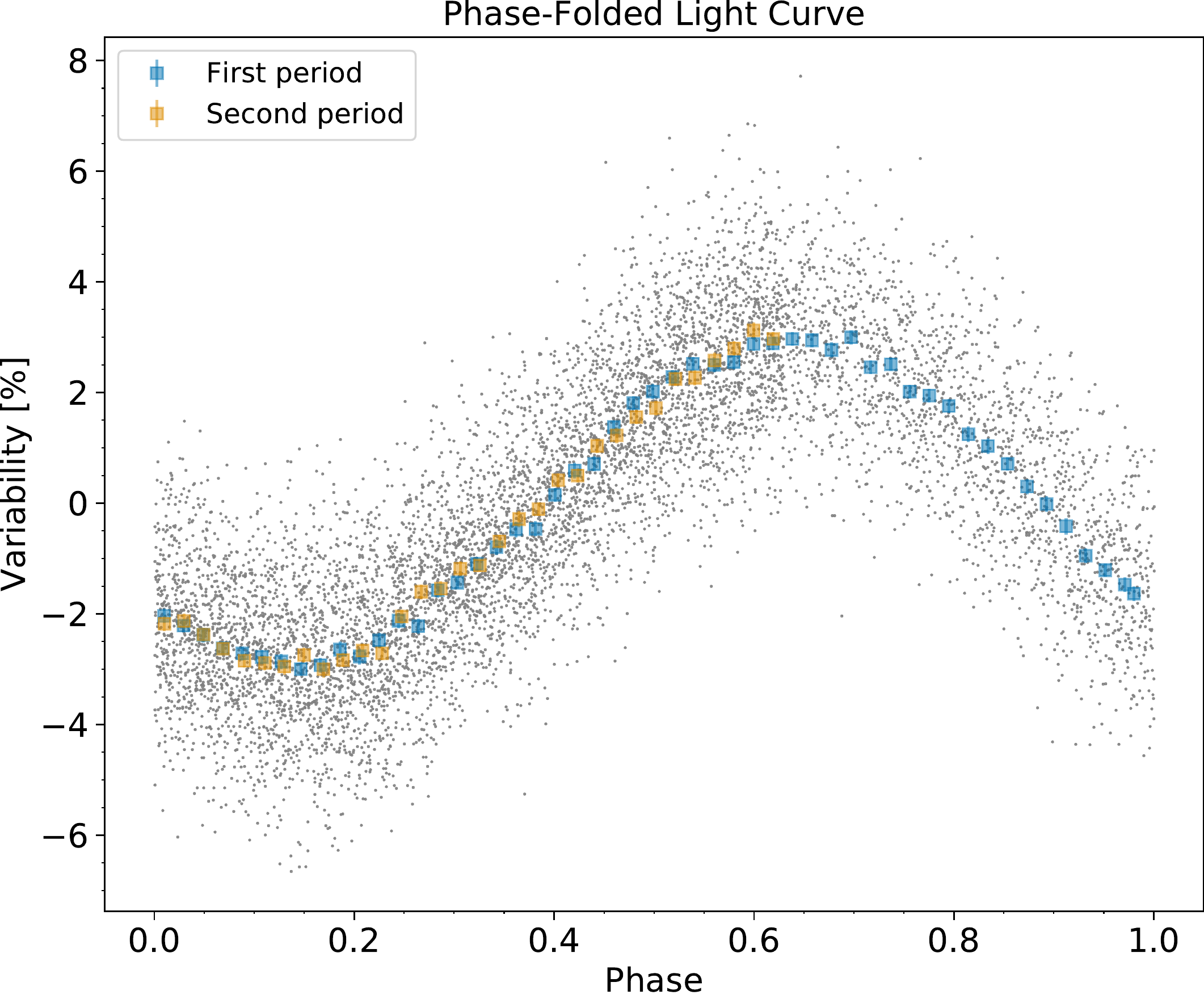}
  \caption{The \emph{Spitzer} light curve of VHS1256b folded to the best-fitting period of 22.04~hr. The blue and yellow squares are binned photometric points in the first and the second period, respectively. The light curves spanning the two periods are in excellent agreement with each other.}
  \label{fig:foldlc}
\end{figure}

We detect strong sinusoidal modulations in the \emph{Spitzer} 4.5 \micron{} light curve of VHS1256b and do not find any persistent variability in the light curve of the host (Figure~\ref{fig:correctedlc}). Assuming a single sine wave model, we find a best-fitting period of $P=22.04\pm0.05$ hr and a peak-to-peak amplitude of $A=5.76\pm0.04$\% with a reduced $\chi^{2}$ of 1.26 (\edit1{DOF=7953}). Figure~\ref{fig:foldlc} shows the light curve phase-folded to the best-fitting period. The 36 hr  observations cover 162.9\% of a full rotation period. Over the phase where our target is observed twice, \edit1{we do not find any evidence that the light curves in two periods differ from each other (two-sample KS test $p>0.99$).}

To explore possible complexities in the light curve, we attempt to fit higher order Fourier series, as adopted in \citet{Metchev2015} and \citet{Yang2016}. We find that an increase of the Fourier series orders has a negligible effect in reducing  the fitting residuals. Because of the increased number of free parameters, Fourier series with more than one wave  (number of waves, $k$)  result in even higher values of Bayesian information criterion \citep[BIC,][]{Schwarz1978} (e.g., $\Delta\mathrm{BIC}=\mathrm{BIC}_{k=2}-\mathrm{BIC}_{k=1}=17.49$) than a single sine wave fit, suggesting a single sine wave is strongly favored. Therefore, we conclude a sine wave with  $P=22.04\pm0.05$ hr and $A=5.76\pm0.04$\% is the \emph{simplest and sufficient} model to describe our 36~hr \emph{Spitzer} 4.5~\micron{} light curve of VHS1256b.

The best-fitting period of the \textit{Spitzer} light curve agrees within $1\sigma$ with that obtained from fitting a single sine-wave to its \textit{HST}/WFC3/G141 light curve (Bowler et al. 2020), even though the \textit{HST} light curve only covered less than 40\% of the $22.04$ hr period. This agreement demonstrates the accuracy of the ``single sine wave''  assumption in the \textit{HST} light curve analysis. The possibility of a double- or multiple-peaked light curve can be excluded in both the \textit{HST} and  the \textit{Spitzer} light curves. Therefore, we interpret the $P=22.04$~hr modulation as the rotation period of VHS1256b.

\subsection{Spectral Variability from 1 to 5 \micron{}}
\label{sec:specvar}
\begin{figure}[!t]
  \centering
  \includegraphics[width=\columnwidth]{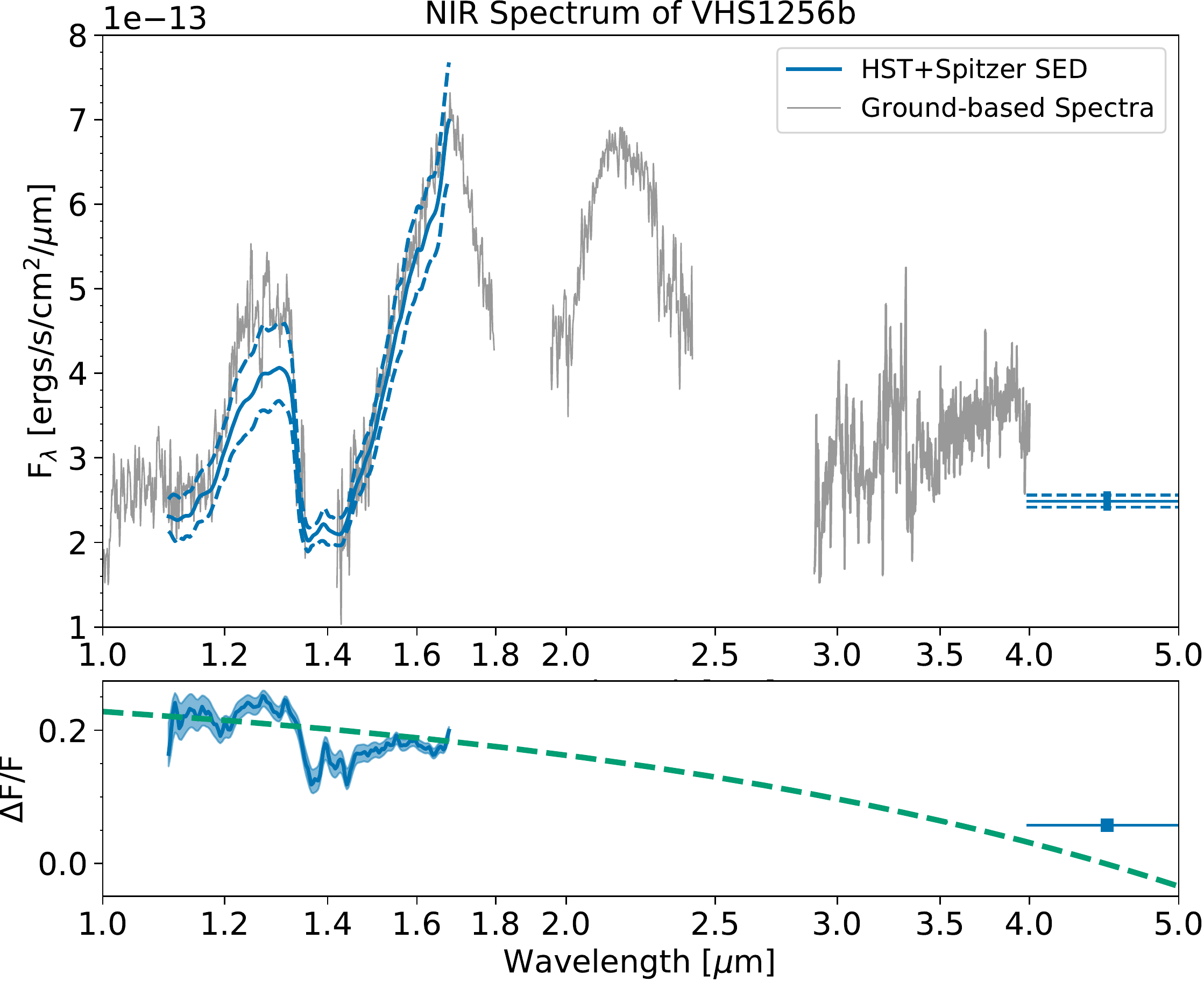}
  \caption{The spectra and spectroscopically-resolved rotational modulations of VHS1256b. In the upper panel, we plot the spectral energy distributions of VHS1256b in its maximum brightness and minimum brightness (dashed lines), and the mean stages (solid line). For comparison, we also show the ground-based spectra that were published in \citet{Gauza2015} and \citet{Miles2018}. The lower panel presents the flux ratio between spectra in two extrema. The green dashed line shows the best-fit linear relationship. The fit does not include the data in the water band. The $x$-axes in both plots are logarithmic.}
  \label{fig:spectrum}
\end{figure}

\begin{figure*}[!th]
  \centering
  \includegraphics[width=\textwidth]{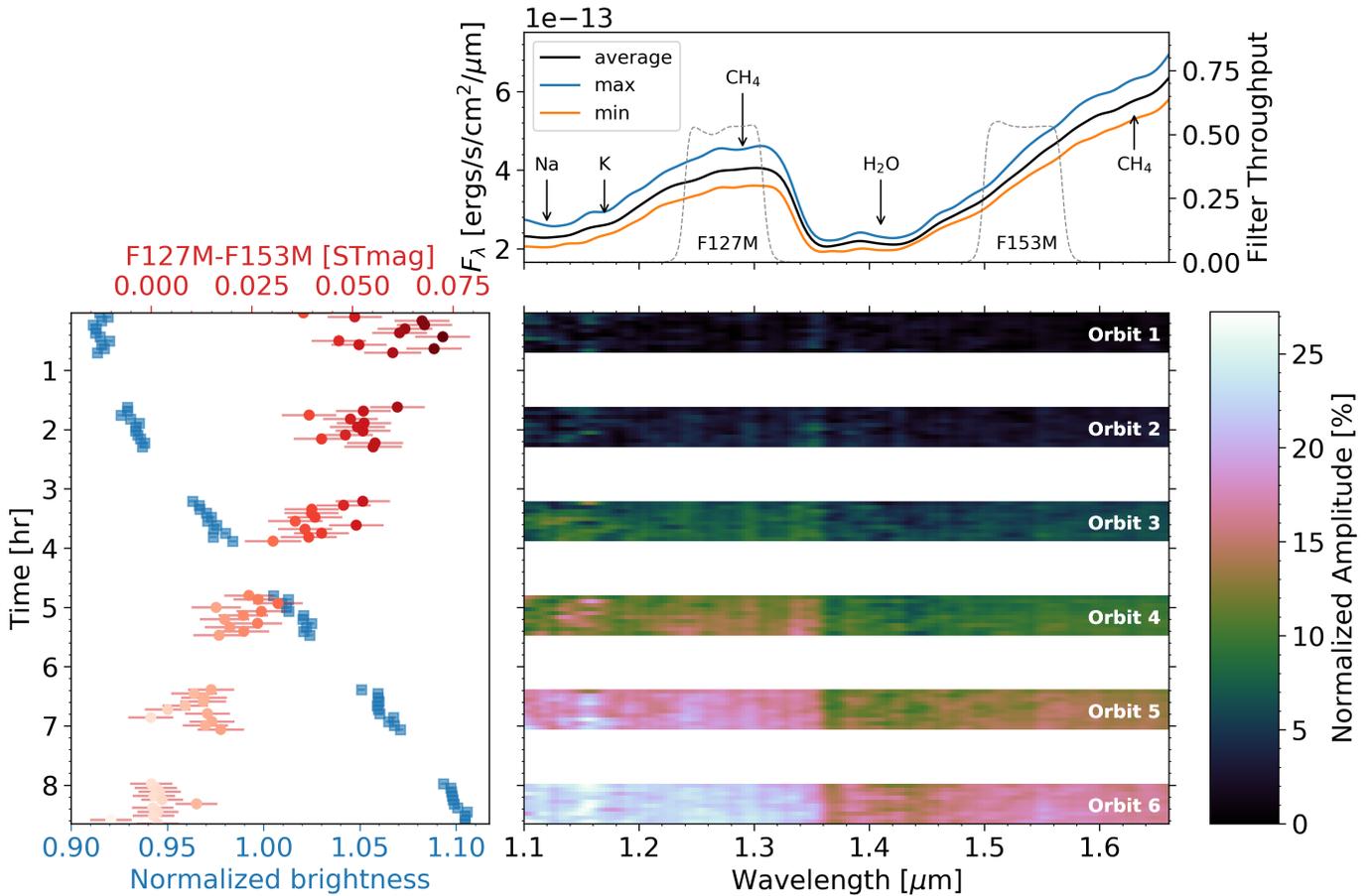}
  \caption{Time-resolved spectroscopic modulations of VHS1256b in the \textit{HST}/WFC3/G141 band (1.1-1.7\micron{}). The lower right panel presents the spectroscopic evolution of VHS1256b. The color represents variability at each wavelength normalized to the faintest state (in the first orbit). Because of the way in which this sequence is normalized, the chromatic variations  are best exhibited at the brightest state, i.e., the end of the time series. The upper  panel shows the maximum, minimum, and the mean spectra. The F127M and F153M filter throughput profiles are also presented. The left  panel shows the integrated flux and the F127M--F153M color variations.}
  \label{fig:specImage}
\end{figure*}

We combine the \emph{Spitzer} flux density and modulation amplitude measurement with those from \textit{HST}/WFC3 and plot the results in Figure~\ref{fig:spectrum}.  For comparison, we also include the ground-based near-infrared spectra from \citet[$J/H/K$ bands]{Gauza2015} and \citet[$L$ band]{Miles2018}. Flux calibrations for the ground-based $J/H/K$-band spectra were conducted for each band individually to match the VHS $J/H/K$-band photometry. The published $L$-band spectrum has already been flux calibrated. The ground-based $J$- and $H$-band spectra and the maximum \textit{HST}/G141 spectrum are in excellent agreement, which suggests that the VHS photometry was taken when VHS1256b was at its maximum brightness.

The modulation amplitude of VHS1256b is clear wavelength dependent. The 1.4\micron{} water absorption band has a lower variability amplitude than its surrounding continua. The amplitude also reduces with increasing wavelength. The amplitude in the \emph{Spitzer} Channel 2 band is 30\% of the average amplitude in the WFC3/G141 band (\edit1{1.1--1.67 \micron{}}), 33\% of the amplitude in the water band (integrated using the WFC3 F139M filter transmission curve), and only 22\% of the amplitude in the F127M band (close to the $J$ band). Excluding the water band, a linear fit to the amplitude-wavelength curve yields a slope of $-6.67\%/\micron{}$.  However, the amplitude-wavelength relation cannot be adequately modeled with a slope. For example, the \emph{Spitzer} Channel 2 amplitude is significantly higher than the best-fit line.

\edit1{We note a caveat in the comparison of modulation amplitudes that are asynchronously observed. The substellar atmosphere and cloud structures can change on the timescales of tens to hundreds of rotations \citep[e.g.,][]{Apai2017,Tan2018}.  In our case, the \emph{Spitzer} and \textit{HST} observations are separated by $\sim{}$557 Earth days, or more than 600 rotations of VHS1256b. The atmosphere of VHS1256b is likely to have undergone substantial evolution over that period. Only when the cloud-covering fraction vertical structures are the same in the \textit{Spitzer} and \textit{HST} observations, do we expect the modulation amplitudes in the two epochs to have the same wavelength dependence and can we meaningfully compare the two sets of measurements (demonstrated in the lower panel of Figure~\ref{fig:spectrum}). This condition is met when the clouds in the two epochs share similar heterogeneity and are formed by the same species with the same sedimentation efficiency and atmospheric thermal profiles.}

In Figure \ref{fig:specImage}, we summarize the WFC3/G141 spectroscopic modulations in a time-resolved manner, as opposed to the comparison between extrema. We seek to condense all relevant measurements, including temporal brightness modulations, near-infrared color changes, and spectral variations in this one presentation. The left panel shows the broadband light curve and the near-infrared color (F127M-F153M) variability. The near-infrared color is presented in $\Delta$STmag ($-2.5\log(f_{\lambda})-21.1$) between synthetic photometry in the F127M and F153M bands. The 1.1--1.7~\micron{} integrated flux and the near-infrared color are anti-correlated: while the brightness increases by $~20\%$, the F127M-F153M color decreases (becomes bluer) by 0.065mag. This result is in agreement with the near-infrared color evolution of several mid-L to L/T transition type brown dwarfs \citep[e.g.,][]{Apai2013, Lew2020}. The lower right panel of Figure~\ref{fig:specImage}, in which the color map represents flux density variations at each wavelength relative to its faintest value, displays the time-resolved spectroscopic variability of VHS1256b during the \textit{HST} observations. In this plot, small-scale structures are also evident in addition to the near-infrared color and water absorption depth variations. Close inspection reveals that the modulation is slightly lower at 1.12, 1.17, 1.28, and 1.63 \micron{}.  These wavelengths overlap with absorption lines or bands of Na, K, and CH$_{4}$.

\subsection{Principal Component Analysis of the \textit{HST}  Time-resolved Spectra}

\begin{figure*}
  \centering
  \plottwo{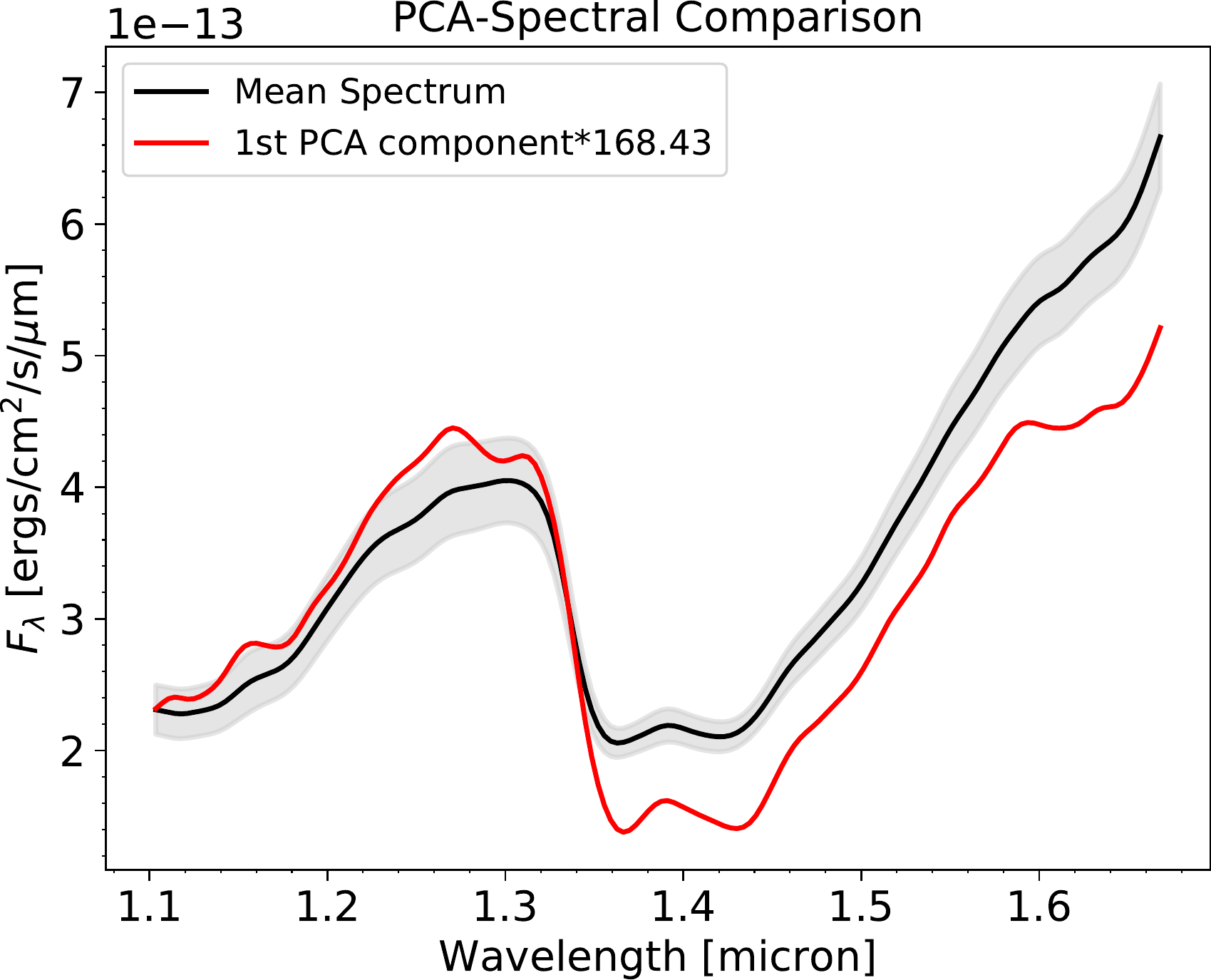}{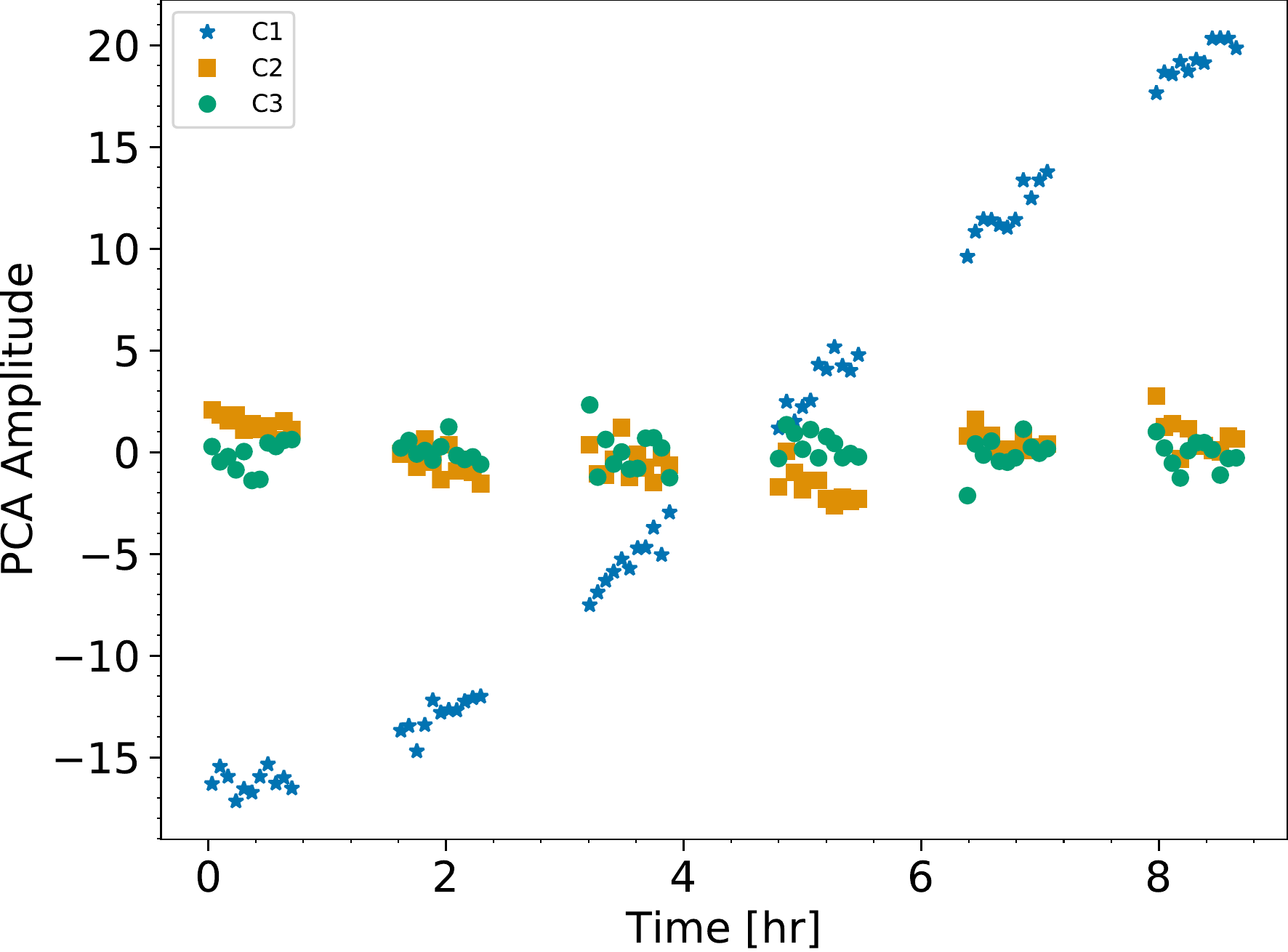}
  \caption{Principal component analysis results for the WFC3/G141  time-series spectra of VHS1256b. The PCA suggests that the first principal component (PC1, left panel) explains the  majority (96.7\%) of the spectral variance. PC1 is scaled up by a factor of 168 so that the average 1.1 to 1.67 \micron{} flux densities in PC1 and the mean spectrum are equal. Compared to the average spectrum, PC1 has a bluer near-infrared color, a deeper water absorption band, and stronger variations in the alkali and methane features. \edit1{However, only the near-infrared color, the water absorption band, and the 1.67 \micron{} methane feature correspond to significant variability amplitude changes in the light curves.} The right panel shows the contributions of the first three PCs to the spectrum as a function of time. The contributions from PC1 are tightly correlated with the overall brightness. The contributions from the second and the third PCs stay relatively constant. The overall brightness and the spectroscopic variability are well explained by  PC1.}
  \label{fig:pca}
\end{figure*}

To quantitatively examine the wavelength dependence of the rotational modulations in Figure~\ref{fig:specImage}, we conduct a principal component analysis (PCA) of the spectral time-series observations. Before the \texttt{PCA} analyses, we first normalize them with the \texttt{StandardScalar} function in the python package \texttt{scikit-learn} so that the series has a mean of zero and a standard deviation of one. The PCA is implemented using the \texttt{PCA} function in \texttt{scikit-learn} with the option ``svd\_solver='full'''.  

Similar to previous PCA studies of other brown dwarf spectral time-series observations \citep[e.g.,][]{Apai2013}, most of the spectral variations can be explained by a single principal component (PC) --- the first PC explains 96.7\% of the variance, and none of the remaining PCs explain more than 1\% of the variance. Therefore, the entire \textit{HST}/WFC3 time-series dataset can be reduced to two dimensions, the mean spectrum (MS) and the first PC (PC1). At time $t$, the spectrum $\mathcal{S}(t)$ of VHS1256b can be expressed as
\begin{equation}
  \label{eq:PCA}
  \mathcal{S}(t) = \mathrm{MS} + c(t)\,\mathrm{PC1}
\end{equation}
in which $c(t)$ is the amplitude of the first PC. Figure~\ref{fig:pca} shows the shape of MS and PC1, and the trend for $c(t)$. During the \textit{HST}/WFC3 observations, $c(t)$ increased with time as a result of the increase in overall brightness of $\mathcal{S}(t)$. Comparing PC1 and MS, we find that PC1 has deeper water absorption and a bluer color. Thus as the contribution from PC1 increases with time, the water absorption of VHS1256b deepens and its color becomes less red. Absorption features at 1.12, 1.17, 1.28, and 1.63 \micron{}, which are weak or invisible in the \textit{HST} spectra of VHS1256b, become prominent in PC1. This means that as the contribution from PC1 to $\mathcal{S}(t)$ increases with time, the absorptions at these wavelengths strengthen. The 1.12 and 1.17 \micron{} features correspond to sodium and potassium lines. The 1.28 and 1.63 \micron{} features both overlap with two CH$_{4}$ absorption bands. \edit1{To evaluate the detection significance of these features, we integrate light curves in the wavelength ranges of these features and their neighboring continua and compare the modulation amplitudes between the spectral feature light curves and the continuum light curves. We find that only the 1.63 \micron{} CH$_{4}$ light curve demonstrates significant modulation amplitude difference from its continuum. Therefore we only claim detection of an amplitude variation in this CH$_{4}$ band, but not in the other spectral features.}
\newcommand{\methane}{CH$_{4}$\xspace}

\subsection{Detection of Variability Amplitude Changes in the \methane Band}
\begin{figure*}[!t]
  \centering
  \plottwo{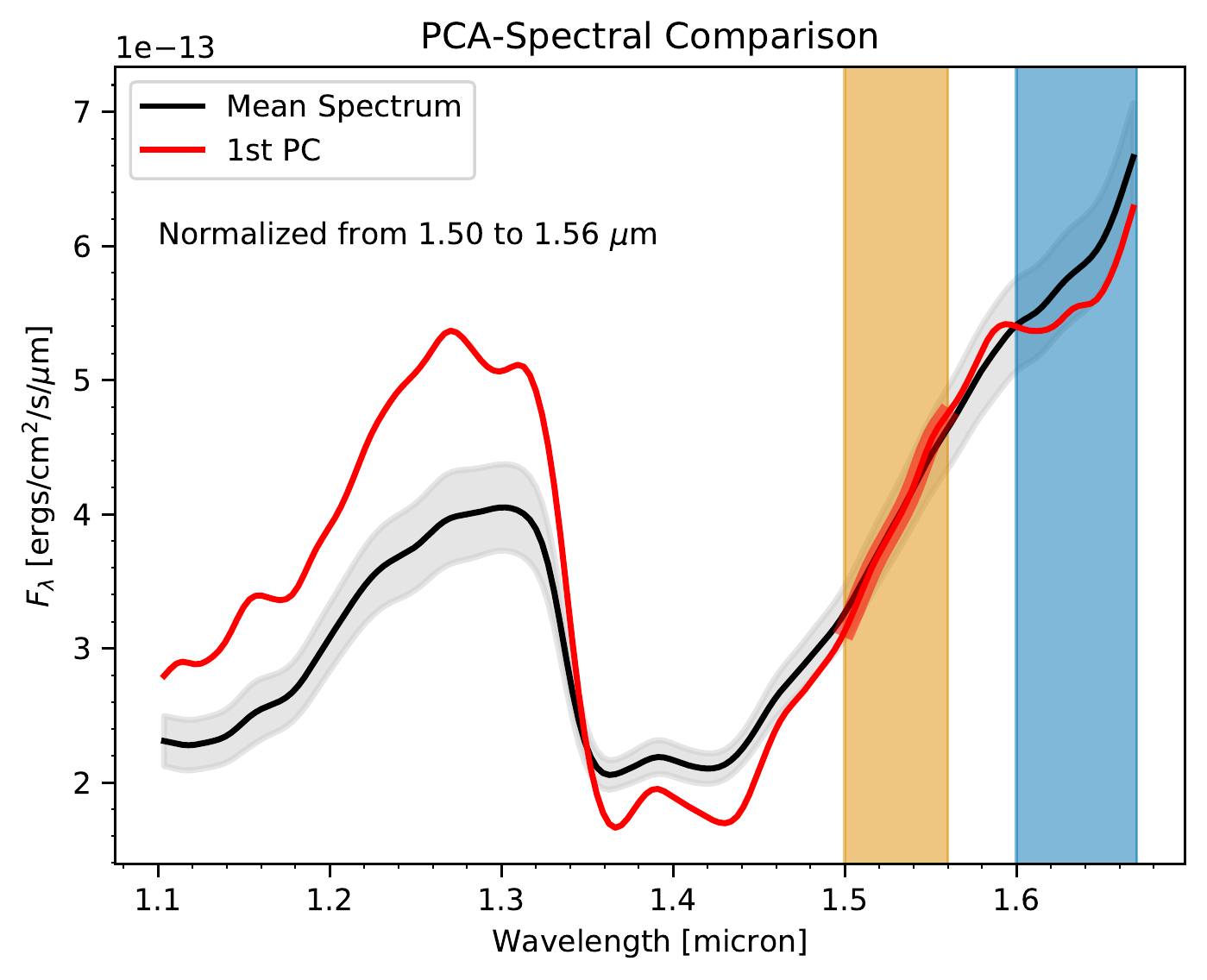}{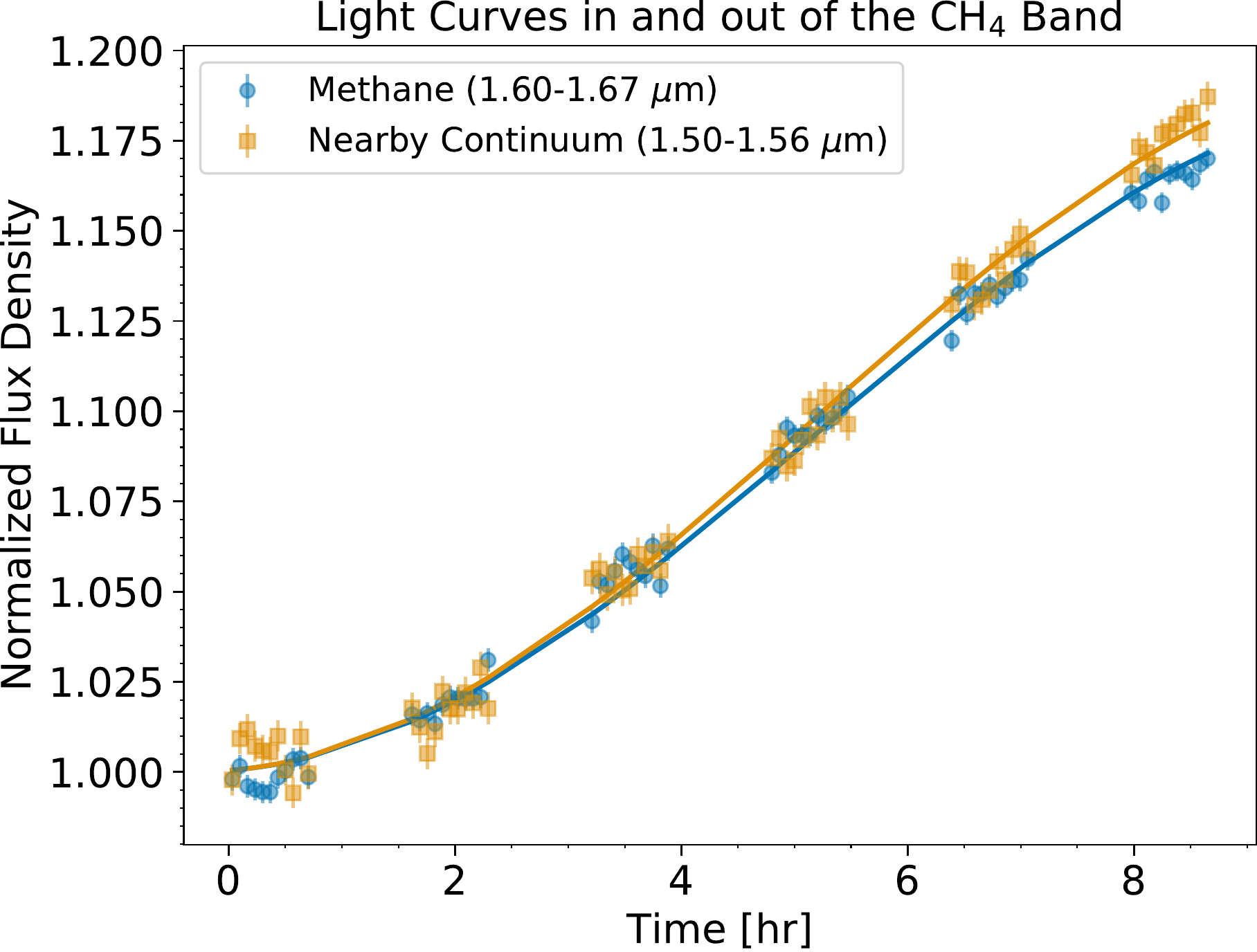}
  \caption{Amplitude variations in the ~1.67\micron{} \methane{} band. The left panel shows the average spectrum and PC1 normalized by their flux density in the 1.50 to 1.56 \micron{} region. A significant absorption feature is present in the PC1 spectrum at wavelengths $>1.60\,\micron{}$. This feature matches the R-branch of the 1.67\micron{} \methane band. The blue (\methane band) and orange (continuum) vertical spans mark the wavelength ranges for the two light curves shown in the right panel.  The two light curves are normalized at their minima, so the difference is most prominent at their maxima, which are at the end of the light curves. The significance of the amplitude difference between these two light curves is $3.28\sigma$.}
  \label{fig:methane}
\end{figure*}

One of the most prominent differences between the average spectrum and PC1 is in the wavelength range of 1.60 to 1.67 \micron{}. The apparent absorption feature in PC1 is clearly present when we normalize the average spectrum and PC1 from 1.50 to 1.56 \micron{}  (Figure \ref{fig:methane}). This feature matches the R-branch of the 1.67\micron{} \methane band. The lower flux density in this region in PC1 compared to the average spectrum therefore indicates the variability amplitude is smaller within the \methane absorption band than in the neighboring continuum.

To confirm this detection, we compare two light curves: one for the \methane band, which is integrated from 1.60 to 1.67~\micron{}, and the other for the continuum, which is integrated from 1.50 to 1.56~\micron{} (Figure~\ref{fig:methane}). To quantitatively evaluate their different modulations, we fit sinusoids to the two light curves and compare the best-fit amplitudes. The periods and the phases of the sine waves are fixed in this analysis. The period is fixed to the one measured from the \textit{Spitzer} light curve (22.04 hr), and the phase is fixed to the best-fit value of the G141 broadband light curve \citep{Bowler2020}. For the light curve in the methane band, the best-fit amplitude (peak-to-peak) is $A_{\mathrm{CH_{4}}}=18.680\pm0.016$\%. For the continuum, the best-fit amplitude is $A_{\mathrm{cont.}}=19.602\pm0.023$\%. The amplitude difference between these two light curves is $\Delta A_{\mathrm{cont.}-\mathrm{CH_{4}}}=0.922\pm0.028$\%, which is $3.29\sigma$ above zero. This indicates a $>3\sigma$ detection of  diminished modulation amplitude within the 1.67\micron{} \methane band compared to the continuum.

\subsection{Atmospheric Models for VHS1256b}
\label{sec:results:model}
\newcommand{\cloudopacity}{\ensuremath{C_{\mathrm{frac}}}\xspace}

We extend our atmospheric modeling work on VHS1256b, which was initially presented in \citet{Bowler2020}, to investigate the physical properties of its atmosphere and clouds following the method presented in \citet{Saumon2008} and \citet{Morley2012}. We calculate atmospheric models assuming radiative-convective and chemical equilibrium varying in effective temperature (\teff), surface gravity (\logg), and  cloud sedimentation efficiency ($f_{\mathrm{sed}}$). For the effective temperature, the model grid includes $\teff=1240$~K (as adopted in \citealt{Miles2018}, which is consistent with the estimate from evolutionary models) and two lower \teff{} values of 1100~K and 1000~K. The surface gravity values include \logg{} of 3.2, 4, and 4.6 (in cgs units). For the sedimentation efficiency, $f_{\mathrm{sed}}$, values of 1.0, 1.5, 2.0, and 2.5 are adopted. Using these self-consistent pressure-temperature profiles, clouds, and atmospheric abundances, we calculate moderate-resolution thermal emission spectra as described in \citet{Morley2015} (see their appendix). We also include models with a reduced cloud optical depth ($\tau_{\rm cloud} = C_{\rm frac} \times \tau_{\rm cloud,\, standard}$) at all pressure levels. This method allows us to model the impact of a reduced number of cloud particles in a localized atmospheric column, which would lead to variability. We keep the cloud scattering properties (single scattering albedo and asymmetric parameter) equal and simply reduce the cloud optical depth. We calculate thermal emission spectra with these reduced cloud optical depths (without recalculating the pressure-temperature profile). 

We also incorporate disequilibrium chemistry in our models. In an atmosphere column where the replenishing rate of CO exceeds the  reaction rate of the CO-to-\methane{} reaction, the mixing ratios of CO and \methane{} deviate from the predictions from models for which chemical equilibrium is assumed. To simulate this process, we assume in our disequilibrium models that the productions of CO, \methane{}, H$_{}2$O, and CO$_{2}$  are quenched above a certain pressure level. Quenching of these molecular species has previously been adopted to investigate disequilibrium chemistry of brown dwarfs and giant exoplanets \citep[e.g.,][]{Barman2011,Barman2015,Moses2016}. For each combination of \teff, \logg{}, $f_{\mathrm{sed}}$, and \cloudopacity, we generate equilibrium models and disequilibrium models with quench levels at 0.19, 2.1, 24.1, and 274 bars.

\begin{figure*}[th]
  \centering
  \includegraphics[width=.46\textwidth]{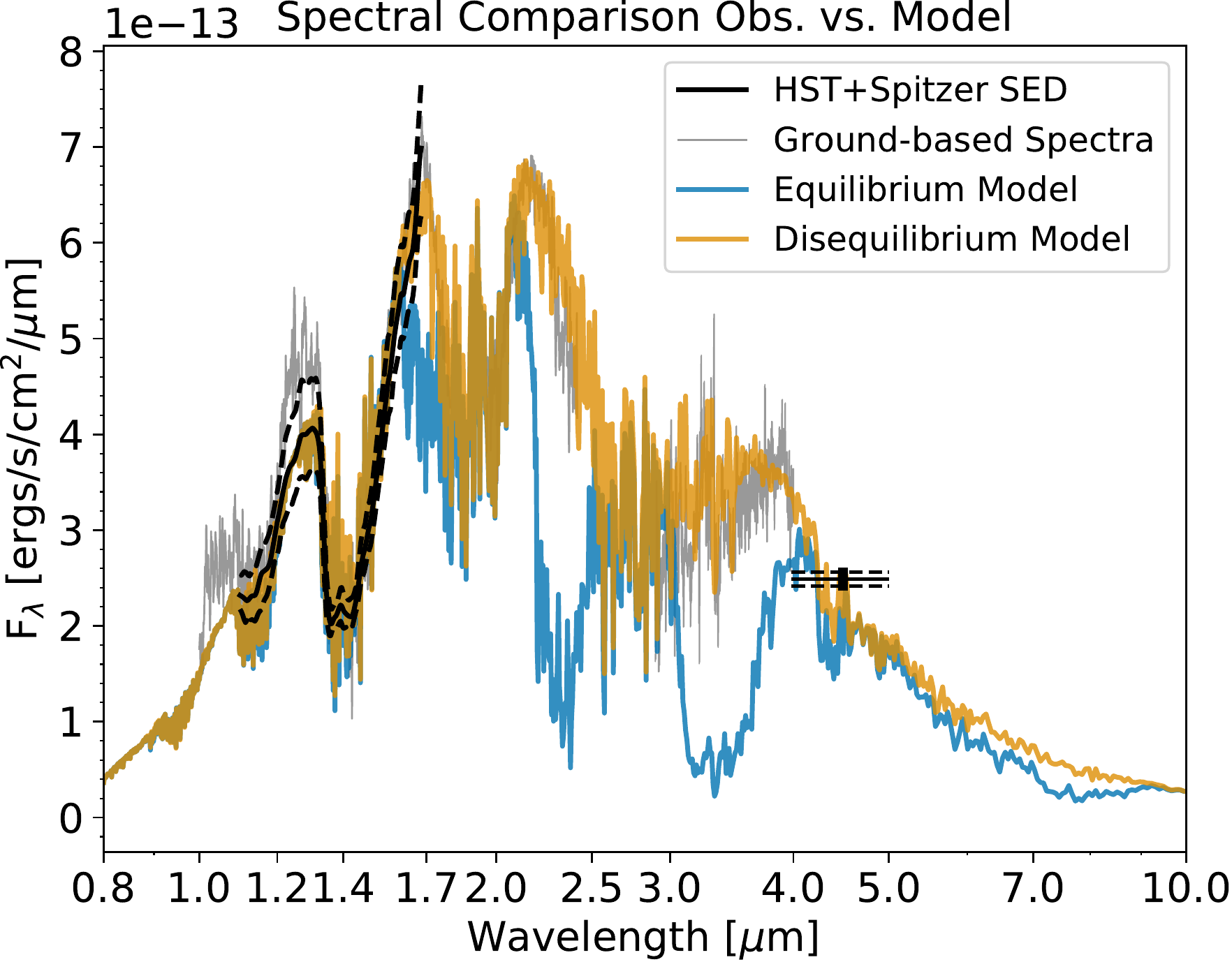}
  \includegraphics[width=.48\textwidth]{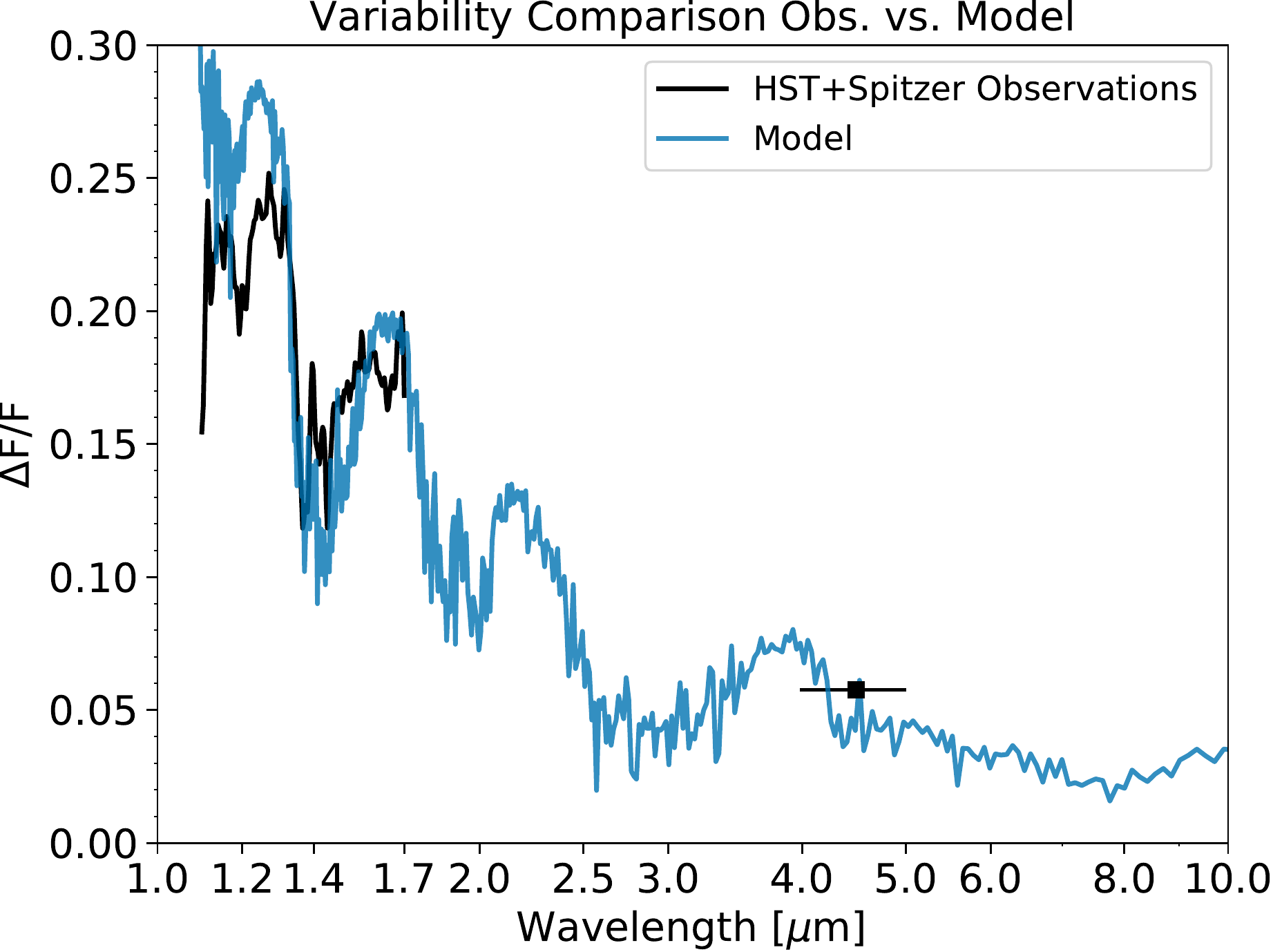}
  \caption{Comparisons between atmospheric models and observations. \textit{Left:} The near-infrared spectrum of VHS1256b is best fit with a model spectrum with \teff=1000~K, \logg=3.2, $f_{\mathrm{sed}}=1.0, \cloudopacity=85\%$, and disequilibrium chemistry for carbon-bearing species. The model spectrum and the observations are in excellent agreement from 1.0--5.0\micron{}. Disequilibrium chemistry is required by the model to reproduce the observed $H$-band spectral slope and the $L$-band flux. \textit{Right:} The observed spectroscopic variability can be well reproduced by perturbing the model cloud opacity. The ratio between  \cloudopacity of 75\% and \cloudopacity of 90\% model spectra matches the observations in the overall 1--5 \micron{} slope and in the variations in the 1.4 \micron{} H$_{2}$O absorption band. The $x$-axes of both plots are logarithmic.}
  \label{fig:model}
\end{figure*}

\edit1{We use both the \textit{HST}+\textit{Spitzer} SED and the ground-based spectra (Figure \ref{fig:model}) to select the best model that minimizes the total squared residuals.} We find that a model with  $\teff=1000$~K, $\logg=3.2$, $f_{\mathrm{sed}}=1.0$, and $\cloudopacity=80\%$ best reproduces the observed spectrum of VHS1256b from 1.0 to 5.0~\micron{}. This result suggests that VHS1256b has a relatively cool temperature, a low surface gravity, and very thick and high clouds. As discussed in \citet{Miles2018}, the $L$-band spectrum of VHS1256b is very sensitive to the chemistry of carbon-bearing species. Our model comparison results strongly favor a disequilibrium chemistry model over an equilibrium one. In the best-fitting model, the CO-\methane{} reaction is quenched at 274 bar. In this model, the \methane{} mixing ratio in the atmosphere above the $\tau=1$ level is several orders of magnitude lower than suggested by the equilibrium. This is in agreement with the findings in \citet{Miles2018}.

To investigate the effect of patchy clouds on spectroscopic variability, we use the models with artificially reduced cloud opacity depths. We find that on the basis of the best-fitting atmospheric model, varying the $\tau_{\mathrm{cloud}}$ by 15\%  can reproduce the observed variability in both the \textit{HST}/G141 and the \textit{Spitzer} Channel 2 bands reasonably well. The ratio between the \cloudopacity of 75\% and \cloudopacity of 90\% model spectra matches the observations in both the 1--5 \micron{} overall trend and the reduced variability in the 1.4 \micron{} H$_{2}$O band. These similarities suggest that inhomogeneous clouds are the dominant source of rotational modulations for VHS1256b. 

\subsection{The Inclination of the Rotational Axis of VHS1256b}

The inclination ($i$) of the rotational axis can be constrained by combining the rotation period measurement with the projected spin velocity ($\vsini$) measurement. With known rotation period ($P$) and radius ($R$), the equatorial rotation velocity ($v_{\mathrm{eq}}$) can be derived as
\begin{equation}
  \label{eq:veq}
  v_{\mathrm{eq}} = \frac{2\pi R}{P}.
\end{equation}
The inclination can then be calculated as
\begin{equation}
  \label{eq:inclination}
  i = \arcsin\Bigl( \frac{\vsini}{v_{\mathrm{eq}}} \Bigr) = \arcsin\Bigl( \frac{\vsini P}{2\pi R} \Bigr)
\end{equation}
\newcommand{\Rjup}{\ensuremath{R_\mathrm{Jup}}}
\newcommand{\Rp}{\ensuremath{R_\mathrm{p}}}

When we assume that VHS1256b has a radius (\Rp) of 1.0 \Rjup{}, the period of 22.04~hr measured in the \textit{Spitzer} light curve corresponds to $v_{\mathrm{eq}}=5.66$~km/s. This is at the lower end of the measured value of \vsini{} of $13.5_{-4.1}^{+3.6}$~km/s from \citet{Bryan2018}. With a significantly larger radius of 2.0 \Rjup{}, the derived $v_{\mathrm{eq}}$ of 11.2~km/s  is still smaller than but formally consistent with the observed \vsini . This suggests that the true \vsini{} value is likely to be lower than the observed value, and $\sin i$ should be very close to one. Therefore VHS1256b is very likely to be viewed equatorially.

\begin{figure}[!t]
  \centering
  \plotone{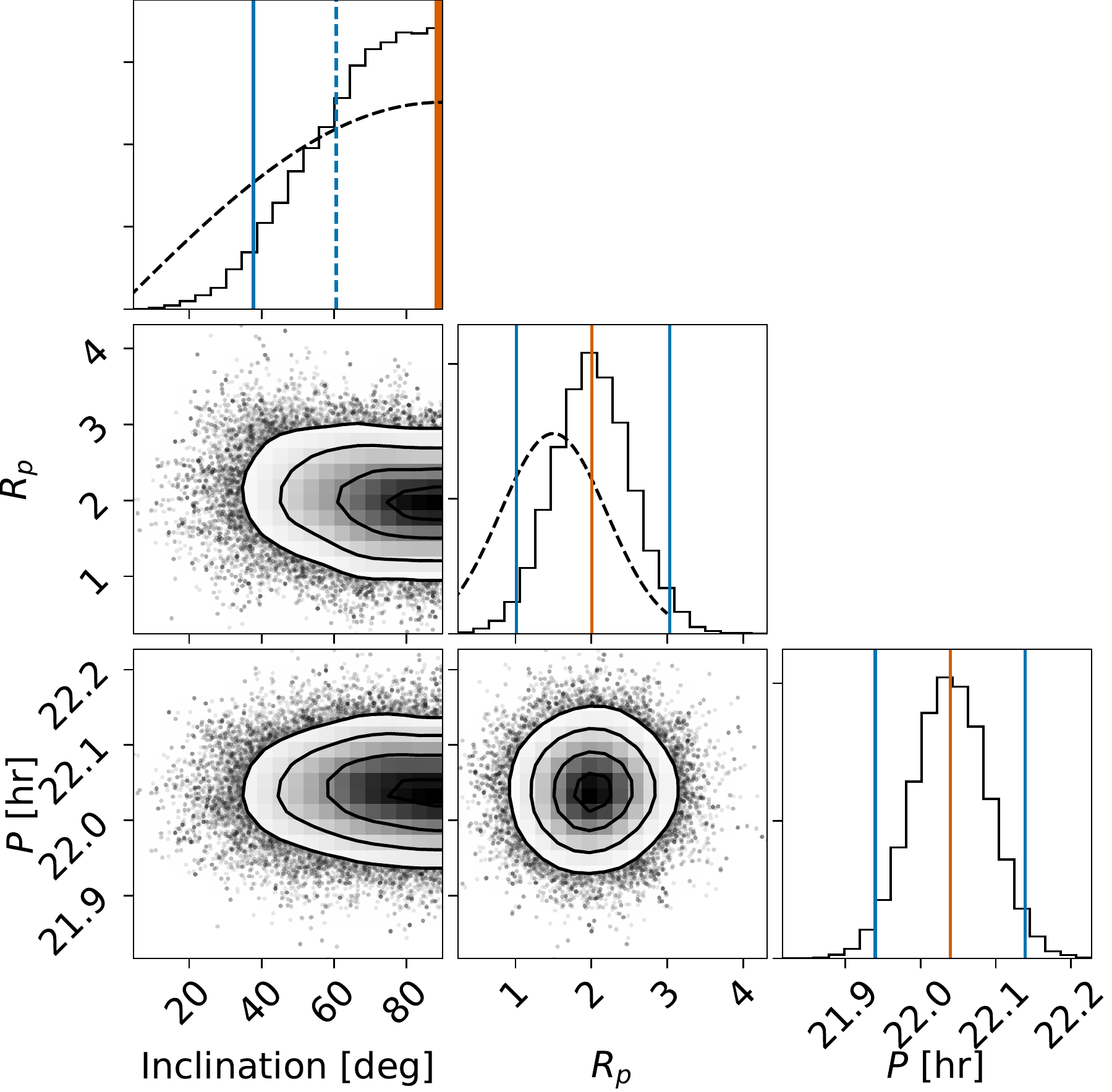}
  \caption{Bayesian analysis results for the inclination of the rotational axis of VHS1256b. In the marginal distribution plots (the top panel in each column),  the dashed black lines are the prior distributions, and the solid black lines are the posterior distributions. The vertical red lines mark the most likely values in the posterior distribution. Solid and dashed vertical lines mark the 68 and 95 percentile ranges. The posterior distribution of the inclination angle favors an edge-on viewing geometry.}
  \label{fig:inclination}
\end{figure}

\newcommand{\Prob}{\ensuremath{\mathrm{Pr}}}
\newcommand{\Pobs}{\ensuremath{P_{\mathrm{obs}}}}
We apply a Bayesian analysis to quantify the probability distribution of the inclination.
Although the radius of VHS1256b is strongly model dependent and only loosely constrained to ${\sim}1$--$2$~\Rjup{} \citep{Rich2016}, with a reasonable assumption of the prior probability distribution of \Rp, we can use Bayes' theorem to derive the joint probably distributions of $(i, \Rp, P)$ of VHS1256b. According to Bayes' theorem,
\begin{equation}
  \Prob(i,\Rp,P|\mathrm{data}) \propto \Prob(i,\Rp,P)\,\Prob(\mathrm{data}|i,\Rp,P),
\end{equation}
in which $\Prob(i,\Rp,P)$ is the prior distribution, $\Prob(\mathrm{data}|i,\Rp,P)$ is the likelihood function, and the data term here includes both the measurements of \vsini and the rotation period (\Pobs). For the prior, we assume that the distributions of $i$, \Rp, and $P$ are independent, thus $\Prob(i,\Rp,P)=\Prob(i)\Prob(\Rp)\Prob(P)$. We assume an isotropic prior for $i$, thus $\Prob(i)=2\int_{0}^{2\pi}\sin(i)\,\mathrm{d}\phi/4\pi = \sin(i)$ \citep[e.g.,][]{Ho2011}\footnote{This is equivalent to $P(\cos i) = 1$. The factor of 2 is from the degeneracy according to which clockwise and counterclockwise rotations are indistinguishable in the observations, and $4\pi$ is the solid angle of the celestial sphere.}. For $\Rp$ we assume a wide Gaussian prior, $\Prob(\Rp)=\mathcal{N}(\mu=1.5\Rjup,\sigma=0.7\Rjup)$, to account for the fact that it is heavily model dependent. As for the period $P$, we adopt a uniform prior from 1 to 50 hr. The likelihood is calculated as
\begin{equation}
  \label{eq:likelihood}
  \begin{split}
    \Prob(\mathrm{data}|i,\Rp,P) =& \frac{1}{\sqrt{2\pi\sigma_{\vsini}^{2}}}\exp\Bigl(-\frac{(\vsini - v_{\mathrm{eq}}\sin i)^{2}}{2\sigma_{\vsini}^{2}}\Bigr) \\
    & \times \frac{1}{\sqrt{2\pi\sigma_{\Pobs}^{2}}}\exp\Bigl(-\frac{(\Pobs - P)^{2}}{2\sigma_{\Pobs}^{2}}\Bigr)
    \end{split}
  \end{equation}
  in which $v_{\mathrm{eq}}$ is from Equation~\ref{eq:veq}, and \vsini, \Pobs, and their uncertainties are from observations. We calculated the posterior distributions using \texttt{emcee} \citep{Foreman-Mackey2012} and present the results in Figure~\ref{fig:inclination}.

  As expected, the Bayesian analysis favors a near-equatorial viewing geometry. The marginal posterior distribution for the inclination angle peaks at $90^{\circ}$. However, its uncertainty is substantial, mostly due to the large uncertainty in the \vsini measurement. The 68 and 95 percentiles of the inclination distribution are $61.5^{\circ}$ and $39.0^{\circ}$, respectively. Additionally, the posterior distribution for the \Rp{} favors a larger radius of $\Rp=2.0\pm0.5\Rjup$ than the average of its prior, which is also nearly $2\sigma$ larger than the \Rp{} value constrained from evolution model. This discrepancy may be caused by inaccurate age assumptions, systematics in the evolutionary models, or biases in the \vsini{} measurements. Future constraints on the system age and the \vsini{} of VHS1256b can help reconcile this mismatch. Nevertheless, this does not change the conclusion that the viewing geometry is equatorial.

\section{Discussion}

\subsection{Clouds in the Atmosphere of VHS1256b}
The 1--5\micron{} variability amplitude observed in VHS1256b is consistent with the interpretation that heterogeneous clouds coupled with rotation are the source of the modulations. Atmospheric models \citep[e.g.,][]{Marley2010,Morley2014} predict that spectral variability induced by patchy clouds is higher in spectral windows formed at deeper layers and at shorter near-infrared wavelengths ($\approx$1--2 $\mu$m) than in molecular absorption bands at longer wavelengths ($\approx$4--5 $\mu$m). Local perturbations of the thermal profiles have the opposite effect \citep{Morley2014,Robinson2014}. Our results agree with the patchy cloud scenario. As demonstrated in \S\ref{sec:results:model}, varying the average cloud optical depth by 15\% in the two hemispheres of VHS1256b reproduces the observed spectral variability. This means that other types of atmospheric heterogeneity (e.g., temperature or molecular abundance variations) should play less significant roles than patchy clouds. The same patchy cloud models also show that a transition from thicker cloud coverage to thinner cloud coverage leads to bluer near-infrared colors and stronger atomic (alkali) and molecular (H$_{2}$O and CH$_{4}$) absorptions in the observations. This matches the fine-scale spectral features observed in our PCA (Figure \ref{fig:pca}) of our \textit{HST}/G141 time-resolved spectra.

An essential caveat in this analysis lies in the fact that our \textit{HST} and \textit{Spitzer} observations were taken in different epochs. Our patchy cloud model that successfully explains the wavelength dependence in the variability amplitude of VHS1256b also predicts that rotational modulations in the \textit{HST}/G141 and the \textit{Spitzer} Channel 2 bands have the same phase. However, without simultaneous observations, we cannot unambiguously test this prediction. Importantly, PSOJ318, whose spectrum is almost identical to VHS1256b, was monitored  by \textit{HST} and \textit{Spitzer} simultaneously and demonstrated a 200--210$^{\circ}$ phase offset between the \textit{HST} and \textit{Spitzer} light curves \citep{Biller2017}. Such a phase offset requires additional variability-inducing mechanisms in the atmospheric models to reproduce the observations.

The sinusoidal shape of the light curves does not support the scenario in which a few spots cause the modulations. Surface maps with only a few large spots result in more irregular light curves instead of sine waves \citep[e.g.,][]{Karalidi2015,Apai2017}.  It is more likely that the atmosphere of VHS1256b is dominated by zonal jet bands and the cloud scale height is modulated by planetary waves. 3D atmospheric dynamic simulations showed that zonal-jets should be ubiquitous for brown dwarfs \citep{Showman2019}. \citet{Zhang2014} introduced two regimes of atmospheric circulation for brown dwarfs: zonal-jet dominated and turbulence dominated. The zonal-jet-dominated atmosphere produces  larger variation amplitudes and can result in sinusoidal rotational modulation on short timescales (a few rotations). The modeling work in \citet{Zhang2014} also showed that the light curve from the jet-dominated atmosphere is likely to evolve on the timescale of tens to hundreds of hours (1 to 10 rotation periods of VHS1256b).  Similar phenomena were observed in long time-baseline monitoring of the variable L/T transition brown dwarfs by \citet{Apai2017}, who interpreted the observations as results of the beating of planetary-scale waves. Additionally, 1D time-dependent atmospheric models demonstrate that radiative cloud feedback can also drive variability on similar timescales \citep{Tan2018}. Our \textit{HST} and \textit{Spitzer} light curves generally agree with each other morphologically, although the observations are separated by several hundred rotations of VHS1256b. The resemblance of the light curve shapes between the two observations  suggests long-lasting zonal features in the atmosphere of VHS1256b, which is similar to the ``possible persistent cloud structure'' in brown dwarf Luhman-16B \citep{Karalidi2016}. Additional monitoring is needed to test whether these atmospheric structures persist in the atmosphere of VHS1256b.

\subsection{CH$_{4}$ in the Atmosphere of VHS1256b}
Our atmospheric models require disequilibrium chemistry to suppress \methane{} absorption features and reproduce the observed flux of VHS1256b at 1.6 \micron{} and in the $L$ band. This is likely a consequence of the low surface gravity of VHS1256b. As illustrated in \citet{Barman2011}, because vertical mixing increases with the decreasing surface gravity, the rate at which convection brings CO from deep in the atmosphere to the photosphere significantly exceeds the chemical reaction rate for converting CO into CH$_{4}$. The upper atmosphere of VHS1256b therefore has more CO and less CH$_{4}$ than the atmosphere in chemical equilibrium. Although the CH$_{4}$ feature was detected in the $L$-band spectrum of VHS1256b \citep{Miles2018}, it is significantly less abundant than the chemical equilibrium value.

Because the production of \methane{} is quenched in our best-fitting atmospheric model, the variability amplitude changes within the \methane{} bands are not as prominent as those within the H$_{2}$O absorption bands in the model spectral variability curve. In our model spectral variability curve (Figure~\ref{fig:model}), the variability amplitude within the 1.60--1.67 \micron{} \methane{} band does not deviate from the continuum amplitude. This is a discrepancy between the observations and our model predictions and calls for additional ingredients in the models in future studies. Thermal-chemical instability that involves transitions from CO to \methane{} has been introduced to replace the role of condensate clouds in explaining the near-infrared colors and spectrum of L-type brown dwarfs with low surface gravity \citep{Tremblin2016,Tremblin2017,Tremblin2019}. Our detection of changes in variability amplitude in the \methane{} band may provide a test for models that involve thermal-chemical instability.

\subsection{Comparison between VHS1256b and Other Variable Very Red Late L-Dwarfs}
\label{sec:discussion:comparison}

\begin{figure*}
  \centering
  \plottwo{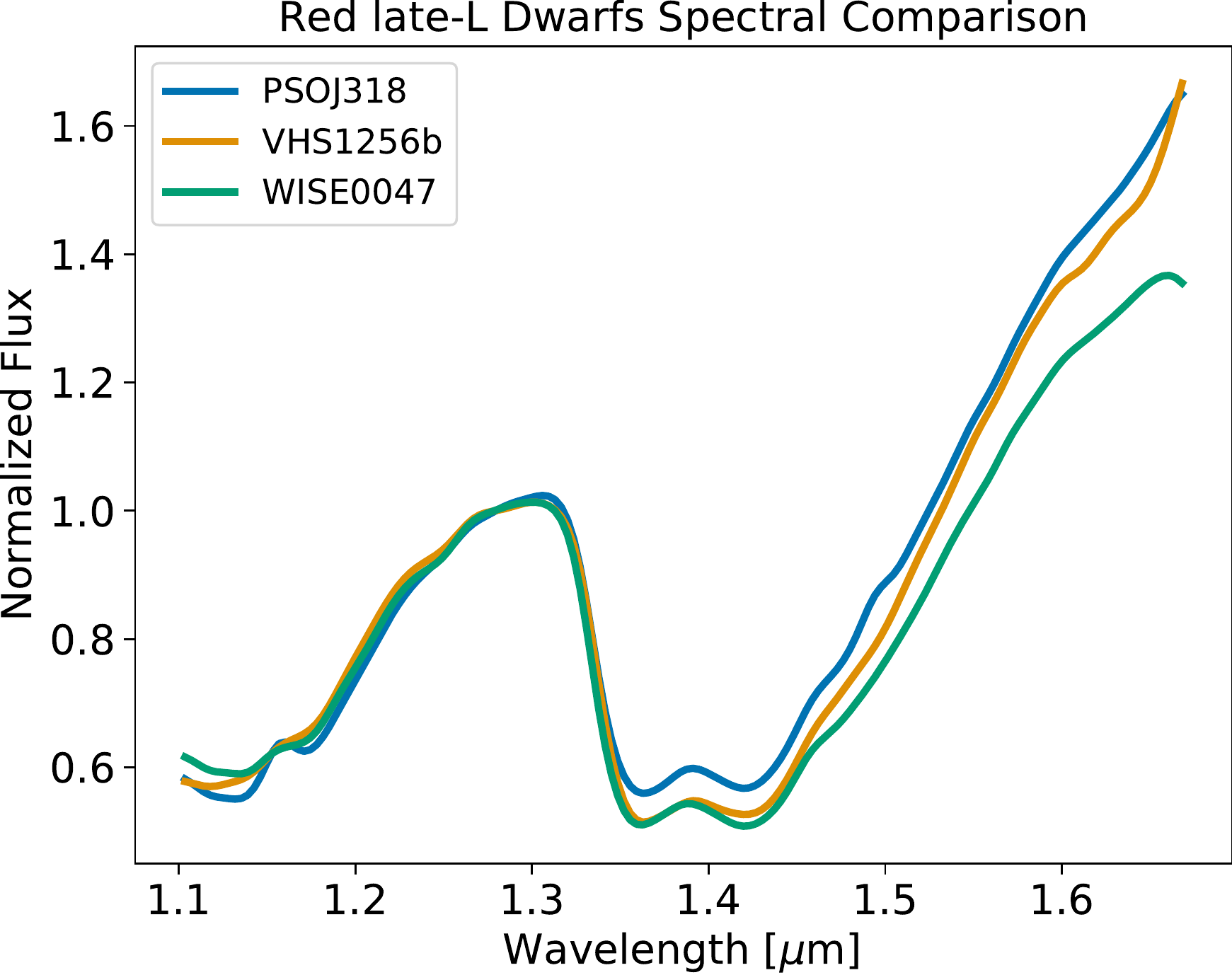}{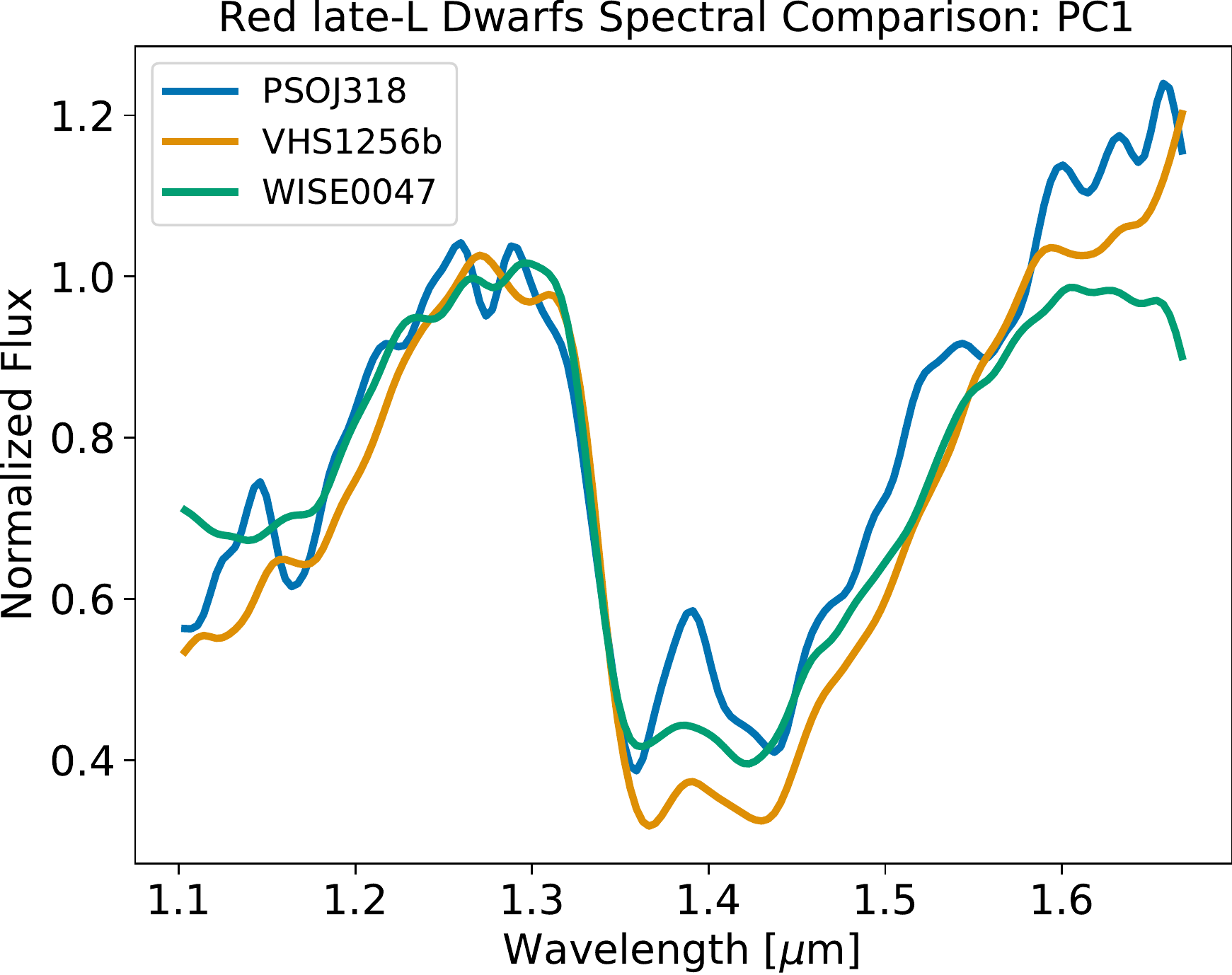}
  \caption{Comparisons of the \textit{HST} spectra and spectroscopic modulations of three late-L variable brown dwarfs: VHS1256b, PSOJ318, and WISE0047. Left: the time-averaged spectra of the three late-L dwarfs. The three spectra are normalized by their flux density at their $J$-band peaks (1.26 to 1.30\micron{}). Right: the first order principal component spectra. The same normalizations as the time-averaged spectra are applied.}
  \label{fig:speccomparison}
\end{figure*}

VHS1256b joins PSOJ318 and WISE0047 as the third late-L type low surface gravity object with a large-amplitude ($>5\%$ peak-to-peak) rotational modulation. These three objects share striking similarities in their near-infrared colors and spectral shapes, and thus in \teff{} and \logg. In Figure~\ref{fig:speccomparison} we compare the \textit{HST} spectra and the spectroscopic variability of VHS1256b, PSOJ318 \citep{Liu2013,Biller2017}, and WISE0047 \citep{Gizis2012,Lew2016}.  PSOJ318 has a shallower water band than the other two objects. WISE0047 has a slightly bluer near-infrared color than VHS1256b and PSOJ318. However, in general, the three spectra are very similar, particularly in the $J$ band.  The close match confirms their nearly identical physical properties  found in previous studies \citep[e.g.,][]{Miles2018}.

More significant differences are evident in the spectroscopic time-dependent components. Both VHS1256b and WISE0047 have significant reductions in variability amplitude within the 1.4~\micron{} water band, but the time-resolved spectra of PSOJ318 do not present a significant detection of this feature. \citet{Biller2017} calculated the spectral ratios between the maximum spectrum and two minimum spectra that were taken at two separate troughs in the \textit{HST} light curves of PSOJ318. They found that the diminished water-band variability amplitude only appeared in one of the ratio curves. This may suggest that PSOJ318 has more complex cloud structures when it was observed than VHS1256b and WISE0047.

We carry out consistent PCA analyses on the time-resolved \textit{HST} spectra of PSOJ318 and WISE0047 as we did for VHS1256b for a direct comparison of the three PC1 spectra (Figure~\ref{fig:speccomparison}). For PSOJ318, we re-reduced the data; for WISE0047, the spectra are from Lew et al. (submitted), which were processed using the same method as the one we applied to our \textit{HST} observations. The PC1 spectrum of VHS1256b has more significant variation at the 1.4\micron{} water absorption band than PSOJ318 and WISE0047. This suggests differences in heterogeneous cloud structures for these three objects. For VHS1256b, cloud heterogeneity is more prominent at the altitude of the $\tau=1$ layer at 1.4\micron{} . For PSOJ318 and WISE0047, however, the clouds are located at higher pressure levels so that the chromatic changes of variability  amplitudes are less pronounced. In addition, amplitude variations in the \methane band are only seen within the PC1 spectrum of VHS1256b. The differences in the first order principal component spectra suggest that for brown dwarfs and planetary-mass objects that have almost identical near-infrared colors and spectra, the spatial and vertical distributions of their clouds can be significantly different and can cause measurable signals in the time-resolved observations. Time-resolved spectroscopy as opposed to a single spectrum provides an especially valuable way to probe substellar atmospheres.



\subsection{The Rotation Rate and Spin-axis Inclination of VHS1256b}
\newcommand{\rjup}{\ensuremath{R_{\mathrm{Jup}}}}
The rotation period of 22.04~hr for VHS1256b is among the longest measured for directly imaged substellar objects. With the assumption of a 2.0~\rjup{} radius, this period is equivalent to an equatorial velocity of 11.32~km/s, or 12.4\% of the break-up velocity (assuming $\logg=4.0$). So far, there has been only one definitive rotation period measurement that is longer than that of VHS1256b \citep[2MASS J16154255 + 4953211, which has a period of 24~hr][]{Metchev2015}, although long-term light curve trends \citep[e.g.,][]{Metchev2015} and small rotational-broadening line profiles \citep{Schwarz2016}  suggest that several brown dwarfs and planetary-mass companions also rotate slowly. The apparent exception in the rotation rate of VHS1256b can be attributed at least in part to observational bias.  Its enormous modulation amplitude prompted the follow-up \textit{Spitzer} observations we presented in this study and led to a determination of its rotation period. Some slowly rotating brown dwarfs and planetary-mass objects with smaller variability amplitude are likely missed in variability programs. 

Alternatively, the typical area of patchy clouds can be proportional to the rotation period based on the Rhines scale argument \citep[e.g.,][]{Apai2013}. The long rotation period of VHS1256b can thus lead to larger cloud structures and stronger rotational modulations than those in faster rotating brown dwarfs. In this case, there may be an underlying population of slowly rotating brown dwarfs that have significant variability amplitudes.

Inclinations of the rotational axes of substellar companions can help trace their formation pathways and dynamical histories. For example, the spin-axis misalignment in the ROX-12 system suggests that the brown dwarf companion either formed through binary-like gravitational fragmentation or was affected by a tertiary body \citep{Bowler2017}. \citet{Bryan2020} found that the planetary-mass companion 2M0122b has a significant spin-orbit misalignment (obliquity). They demonstrated that turbulent motion in the circumstellar disk randomizing the spin-axis orientation is the most favored explanation. For the VHS1256 system, angular momentum elements include the spin and orbit axes of the host binary and the companion. Examining the alignment of these angular momentum vectors will offer valuable insight into the formation of such a hierarchical triple system.


\subsection{VHS1256b as a \textit{JWST} Target}
VHS1256b is a \textit{James Webb Space Telescope} (\textit{JWST}) early release science (ERS) program target (Program ID: 1386, High Contrast Imaging of Exoplanets and Exoplanetary Systems with \textit{JWST}). This program will collect a medium resolution spectrum of VHS1256b from 0.7 to 28.0 \micron{}, which allows precise molecular abundance constraints and the estimates of the carbon-to-oxygen ratio. Our finding that VHS1256b is spectroscopically variable from 1 to 5 \micron{} adds essential astrophysical context for the \textit{JWST} ERS observations. The fact that VHS1256b has strong rotational modulations and is viewed equatorially should be incorporated for interpreting its \textit{JWST} spectrum.


\section{Conclusions}
We present new \textit{Spitzer} Channel 2 time-resolved observations of the  L7 substellar companion VHS1256b. We combine our \textit{Spitzer} light curve with \textit{HST}/WFC3/G141 spectroscopic time-series from \citet{Bowler2020} to study the wavelength-dependent variability of VHS1256b. We summarize our conclusions as follows:

\begin{enumerate}[leftmargin=*]
\item In our consecutive 36~hr \textit{Spitzer} Channel 2 observations, VHS1256b demonstrates strong rotational modulations that are best fit with a sine curve with a period of $P=22.04\pm0.05$~hr, and a peak-to-peak amplitude of $A=5.76\pm0.04$\%. The companion's light curve does not show evidence for evolution or higher complexity than a single sine wave. The light curve of the  host binary is generally consistent with a flat line. We interpret VHS1256b's light curve period as the rotation period of VHS1256b. This is the second-longest rotation period measured for a substellar object, after 2M1615 \citep{Metchev2015}. Its \textit{Spitzer} Channel 2 amplitude is the third-largest observed for an ultracool dwarf, after two T2-type dwarfs 2M2139 and 2M1324 \citep{Yang2016,Apai2017}. The extreme modulations of VHS1256b make it  a prime target for investigating the heterogeneous clouds and atmospheric structures in brown dwarfs and planetary-mass companions.
\item The variability amplitude of VHS1256b exhibits a strong wavelength dependence. The $5.76\%$ amplitude in the 4.5 \micron{} \textit{Spitzer} Channel 2 band is only $30\%$ of the average amplitude observed in the 1.10-1.68 \micron{} \textit{HST}/WFC3/G141 band and $22\%$ of that in the $J$ band. The wavelength dependence of rotational modulations is also apparent in the G141 time-series spectra.  Principal component analysis reveals that the modulations of VHS1256b are not only reduced in the 1.4 \micron{} water absorption band, but also in the alkali lines and the 1.67 \micron{} CH$_{4}$ band.
\item The amplitude difference between the 1.67 \micron{} CH$_{4}$ band and its nearby continuum is detected at $3.3\sigma$ significance. This is the first detection of amplitude variations within the CH$_{4}$ bands for a variable L dwarf. The smaller  amplitude within spectral absorption features and the \textit{Spitzer} Channel 2 bandpass are consistent with the interpretation that heterogeneous clouds cause the rotational modulations.
\item The 1--5 \micron{} SED of VHS1256b is consistent with a model spectrum of \teff of 1000~K, \logg{} of 3.2, and $f_{\mathrm{sed}}$ of 1.0. Disequilibrium chemistry is required in the model to fit the observed flux at 1.6\micron{} and in the $L$-band.
\item The spectroscopic variability of VHS1256b can be well reproduced by varying the cloud optical depth in the atmospheric models. The ratio between two model spectra that have the same \teff, \logg, and $f_{\mathrm{sed}}$ but differ in cloud opacity matches the spectrally resolved variability amplitude from 1 to 5\micron{}. This confirms that heterogeneous clouds are the dominate source of the rotational modulations of VHS1256b.
\item VHS1256b, together with WISE0047 and PSOJ318, forms a group of late-L type dwarfs that have extremely high-amplitude rotational modulations. They share striking similarities in near-infrared colors and spectra, as well as derived physical properties such as \teff{} and \logg. However, they also demonstrate differences in their spectroscopic variabilities. These deviations highlight the advantages of time-resolved observations for characterizing these objects.
\item By combining our rotation period measurement with the $v \sin i $ measurement from high-resolution spectroscopy \citep{Bryan2018}, we find that the rotational axis of VHS1256b prefers a perpendicular orientation to the line of sight. The equatorial viewing geometry reinforces the correlation between variability amplitude and viewing angle found by \citet{Vos2017}. 
\end{enumerate}
\facility{\textit{Hubble Space Telescope}/Wide Field Camera3, \textit{Spitzer Space Telescope}/InfraRed Array Camera}

\software{Numpy\&Scipy \citep{VanderWalt2011}, Matplotlib
  \citep{Hunter2007},  Astropy\citep{Robitaille2013}, Seaborn \citep{Waskom2017},
  pysynphot \citep{Pysynphot}, scikit-learn \citep{scikit-learn}, emcee \citep{Foreman-Mackey2012}, lmfit \citep{Newville2014}}

\acknowledgements We thank the anonymous referee for a constructive referee report. We thank T. Dupuy and M. Liu for reviewing the manuscript and providing constructive comments. We thank B. Gauza and B. Miles for providing the ground-based spectra of VHS1256b, B. W. P. Lew for the \textit{HST} time-resolved spectra of WISE0047, T. Dupuy, M. Liu, and W. Best for helpful communications on the new astrometric and parallax measurements of VHS1256b. This work is based in part on observations made with the \textit{Spitzer Space Telescope}, which is operated by the Jet Propulsion Laboratory, California Institute of Technology under a contract with NASA. This research is based in part on observations made with the NASA/ESA \textit{Hubble Space Telescope} obtained from the Space Telescope Science Institute, which is operated by the Association of Universities for Research in Astronomy, Inc., under NASA contract NAS 5-26555. These observations are associated with program 15197. B.P.B. acknowledges support from the National Science Foundation grant AST-1909209.

\end{document}